\documentclass{aa}  

\usepackage{graphicx}
\usepackage{txfonts}
\usepackage{multirow}
\usepackage{wasysym}
\usepackage[b]{esvect}
\usepackage{natbib}
\begin{document}

   \title{Passband reconstruction from photometry}

   \author{M. Weiler
          %\inst{1}
          \and
          C. Jordi%\inst{1}
          \and
          C. Fabricius
          \and
          J. M. Carrasco%\fnmsep
          }

   \institute{Departament de F{\'i}sica Qu{\`a}ntica i Astrof{\'i}sica, Institut de Ci{\`e}ncies del Cosmos (ICCUB), Universitat de Barcelona (IEEC-UB), Mart{\'i} i Franqu{\`e}s 1, E08028 Barcelona, Spain\\
              \email{mweiler@fqa.ub.edu}
             }

   \date{Received 19 December 2017; accepted 01 Feburary 2018}

% \abstract{}{}{}{}{} 
% 5 {} token are mandatory
 
  \abstract
   {Based on an initial expectation from laboratory measurements or instrument simulations, photometric passbands are usually subject to refinements. These refinements use photometric observations of astronomical sources with known spectral energy distribution.}
   {This work investigates the methods for and limitations in determining passbands from photometric observations. A simple general formalism for passband determinations from photometric measurements is derived. The results are applied to the passbands of HIPPARCOS, Tycho, and {\it Gaia} DR1.}
   {The problem of passband determination is formulated in a basic functional analytic framework. For the solution of the resulting equations, functional principal component analysis is applied.}
   {We find that, given a set of calibration sources, the passband can be described with respect to the set of calibration sources as the sum of two functions, one which is uniquely determined by the set of calibration sources, and one which is entirely unconstrained. The constrained components for the HIPPARCOS, Tycho, and {\it Gaia} DR1 passbands are determined, and the unconstrained components are estimated.}
   {}


   \maketitle
%
%-------------------------------------------------------------------

\section{Introduction}

The full exploitation of photometric data sets requires the knowledge of the passbands in which the photometric measurement have been performed. These passbands result from a combination of different wavelength-dependent instrumental effects, such as the quantum efficiencies of the detectors, the reflectivity or transmissivity of optical elements in the telescope and instrument, and the transmissivities of employed filters. From simulations and laboratory measurements on individual components, a prediction of the passband is usually possible. In operation however, differences between the true passband and the expected one may occur. Such differences may be time-dependent, resulting from ageing, contamination of optical surfaces, and radiation effects (in particular for space-borne instruments). For ground-based observations, the Earth's atmosphere also affects the passband. A re-evaluation of the passband therefore is desirable, based on the photometric measurement of astronomical objects with known spectral energy distributions (SEDs in the following).\par
A frequently used approach in updating the passband with respect to the initial expectation is applying some modifications to the initial passband estimate, such that a better agreement of synthetic photometry and observed photometry for sources with known SEDs is achieved. An early approach in refining passbands is the use of linear combinations of different passbands \citep{Johnson1952}. In this work, we follow a more systematic approach to the problem of passband determination. First, we investigate the principal possibilities and restrictions in constraining a passband from photometric observations of sources with known SEDs (the {\it calibration sources}). Based on the results, we can derive criteria for an optimal selection of calibration sources, and answer the question of how to optimally modify the initial passband. Furthermore, we discuss the effects of uncertainties in the observational photometry and in the SEDs of the calibration sources.\par
We apply the theoretical results derived in this work to four different space-based photometric systems. These are the HIPPARCOS  \citep{Perryman1997} and two Tycho-2 \citep{Hoeg2000} datasets, and the {\it Gaia} data release 1 ({\it Gaia} DR~1 in the following, \citet{Gaia2016a}). These photometric data sets are interesting for both their large contents and their accuracy. The HIPPARCOS catalogue \citep{ESA1997} comprises more than $10^5$ sources, while the Tycho-2 catalogue, also derived from the HIPPARCOS mission, includes about $2.5\times 10^6$ objects. The first data release of the {\it Gaia} mission contains over $10^9$ sources \citep{Gaia2016b}. Taking benefit from stable observing conditions in space, these missions reach photometric accuracies down to the milli-magnitude level in one passband for the HIPPARCOS catalogue (labeled $H_p$ in the following), somewhat lower accuracies  in two passbands for the Tycho-2 catalogue (labeled $B_T$ and $V_T$ in the following), and again a higher accuracy in one passband for the {\it Gaia} DR1 (labeled $G$ in the following).\par
For HIPPARCOS and Tycho, passband estimates have been provided by \citet{ESA1997}, already improved from on-ground calibrations using observations of standard stars. Further re-calibrations have been provided in the past \citep{Bessell2000, BessellMurphy}. For {\it Gaia} DR1, \citet{Jordi2010} provide a pre-launch expectation for the $G$ passband, which has been improved by \citet{Maiz2017}.\par
In this work, we formulate the problem in a basic functional analytic framework in Sec. \ref{sec:formulation}. A functional analytic approach has already been applied in photometry to the problem of photometric transformations and passband design by \citet{Young1994}. It was demonstrated there that, in exchange for the somewhat higher level of abstraction, one obtains an elegant mathematical formulation and deeper insights into the photometric problems. By using a functional analytic approach in the problem of passband reconstruction, we can take benefit of the vector properties of the functions involved (i.e., the SEDs and passbands), and derive simple results on the passbands, the limits in determining the passbands, and criteria for selecting optimal calibration sources (Sec. \ref{sec:unconstrained}). For making the results suitable for practical purposes, we introduce functional principal component techniques to the SEDs (Sec. \ref{sec:basis}) and the corresponding propagation of uncertainties (Sec. \ref{sec:uncertainties}). Equipped with these tools, we then determine the passbands for the HIPPARCOS photometry, the Tycho $B_T$ and $V_T$ photometry, and the {\it Gaia} DR 1 photometry in Sec. \ref{sec:passbands}. We then compare our results with previously published passbands (Sec. \ref{sec:compWithOthers}), discuss practical means for handling systematic errors resulting from the choice of the calibration sources (Sec. \ref{sec:orteffect}), and derive the zero points of the passbands (Sec. \ref{sec:zeropoints}). Finally, we discuss the choice of wavelength resolution for practical purposes (Sec. \ref{sec:wavelengthResolution}), before closing this work with a summary and discussion (Sec. \ref{sec:Summary}).

\section{Formulation of the problem \label{sec:formulation}}

In this work, we perform all computations in terms of photon counts. The spectral energy distribution we specify in terms of photons per unit of time, area, and wavelength. Thus, we better speak of the {\it spectral photon distribution} (SPD in the following), rather than of the spectral energy distribution. The spectral energy distribution can be converted into the SPD, denoted $s(\lambda)$, in units of photons per units of time, area, and wavelength by multiplication with the energy of a photon, that is $h\, c / \lambda$, $h$ being the Planck constant and $c$ the speed of light. We understand the passband as the fraction of incoming photons detected by the instrument, as a function of wavelength. This definition corresponds to what is usually called the {\it photon response curve}, which may be normalised with respect to its maximum value. We take the photometric observation not specified as a magnitude, but as a counting rate, i.e. photons (or, rather, photo-electrons) per unit of time and area. Such an approach is closer to the actual measured quantity of a photon counting detector such as a CCD, and simplifies the computations in this work. For the transformation between photon count rates and magnitudes $m$ we assume a relation
\begin{equation}
m = -2.5 \cdot {\rm log}\left(\bar{c}\right) + zp \quad , \label{eq:magnitude}
\end{equation}
with $\bar{c}$ being the weighted mean photon counts per unit of time and area in a passband $p(\lambda)$, and $zp$ the zero point for the passband used. $\bar{c}$ is related to the passband via
\begin{equation}
\bar{c} = \frac{\int_0^\infty\, p(\lambda) \cdot s(\lambda) \, {\rm d}\lambda}{\int_0^\infty\, p(\lambda) \, {\rm d}\lambda} \quad . \label{eq:i1}
\end{equation}
As the denominator in this equation can as well be absorbed in the zero point, its value is actually not of relevance in this definition. We note that the definition of the magnitude according to Eq. (\ref{eq:magnitude}) is not equivalent to the definition of the magnitude in terms of mean energy flux within the passband, $\bar{f}$. For $\bar{f}$ one obtains
\begin{equation}
\bar{f} = \frac{\int_0^\infty\, p(\lambda) \cdot s(\lambda) \cdot hc/\lambda \, {\rm d}\lambda}{\int_0^\infty\, p(\lambda) \, {\rm d}\lambda} \quad , \label{eq:i2}
\end{equation}
and there can be no general relationship holding between the integral expressions in the numerators of Eqs. (\ref{eq:i1}) and (\ref{eq:i2})\footnote{This is sometimes stated otherwise, e.g. in Eq. (A13) of \citet{BessellMurphy}. There, the reason is the erroneous appearance of the function $S^\prime=R(\lambda)\sigma(\lambda)$ instead of $R(\lambda)\eta(\lambda)$ in Eq. (A11).}. After these preliminaries, we proceed to the problem of passband determination.\par
Let $p(\lambda)$ be some passband. We consider a wavelength interval $I=[\lambda_0,\lambda_1]$, such that we can a-priori assume $p(\lambda)$ to be identical to zero everywhere outside the interval $I$. Assume that for $N$ astronomical sources, photometric observations (i.e. values for the number of photons per unit of time and area) $c_i$, $i=1,\ldots,N$, in the passband $p(\lambda)$ are available, as well as the SPDs of the $N$ sources over the interval $I$, denoted $s_i(\lambda), \; i=1,\ldots,N$. We thus obtain
\begin{equation}
c_i = \int\limits_{\lambda_0}^{\lambda_1}\, p(\lambda)\, s_i(\lambda)\, {\rm d}\lambda\; , \quad i=1,\ldots,N\; . \label{eq:1}
\end{equation}
We are concerned in determining $p(\lambda)$ from the set of $N$ calibration sources. Now we observe that the spectral energy distributions $s_i(\lambda)$, as well as the passband $p(\lambda)$, are square integrable functions over the interval $I$, i.e.
\begin{equation}
\int\limits_{\lambda_0}^{\lambda_1}\, \left[\, s_i(\lambda) \,\right]^2\, {\rm d}\lambda < \infty \label{eq:2}
\end{equation}
for any source $i$, and an analogous expression holds for $p(\lambda)$. Actually, SPDs and passbands have to satisfy even stronger requirements than square integrability, as for physical reasons, these functions have to be non-negative and bound. The square integrability however allows us to make use of the fact that the set of all square integrable functions on the interval $I$ form a vector space with an inner product, which is the Hilbert space ${\mathcal L}^2(I)$ over the field of real numbers. The inner product of this vector space is given by
\begin{equation}
\langle\, s_k \,  | \,  s_l \, \rangle := \int\limits_{\lambda_0}^{\lambda_1} \, s_k(\lambda) \cdot s_l(\lambda) \, {\rm d}\lambda \quad , \label{eq:3}
\end{equation}
$s_k(\lambda), \; s_l(\lambda) \in {\mathcal L}^2(I)$. The reader not familiar with the concept of considering functions as vectors in a vector space of functions may find it useful to think of it in the same way as of the familiar vectors in an Euclidian vector space. This analogy is justified by the fact that Euclidian spaces, such as, say ${\mathbb R}^3$, are also realisations of Hilbert spaces, differing from the Hilbert spaces of functions mainly in the fact that the later are infinite dimensional, while the more ''typical'' vector spaces have finite dimensionality. A wide reaching one-to-one correspondence between Euclidian spaces and the Hilbert space of square-integrable functions is obtained by replacing the ''typical'' vectors of Euclidian space by square-integrable functions, and the standard dot product of Euclidian space, $\vv{r}_1 \cdot \vv{r}_2$ by the expression given by eq. (\ref{eq:3}). Then, just as a vector in ${\mathbb R}^3$ can be developed as a sum of basis vectors, $\vv{r} = x\cdot \vv{e}_1 + y \cdot \vv{e}_2 + z \cdot \vv{e}_3$, a function $f(\lambda)$ can be developed in a (infinite) sum of basis functions, $f(\lambda) = \sum_{i=0}^\infty \, a_i \cdot \varphi_i(\lambda)$. The basis vectors of ${\mathbb R}^3$ are conveniently chosen to be orthonormal, i.e. satisfying the condition $\vv{e}_i \cdot \vv{e}_j = \delta_{i,j}$. In the same way, the basis functions may for convenience by chosen orthonormal, by satisfying the corresponding condition $\langle\, \varphi_i \, | \, \varphi_j \, \rangle = \delta_{i,j}$. The length of a vector in ${\mathbb R}^3$ (or, more precisely, its $l_2-$norm) is given by $||\, \vv{r}\, ||_2 = \sqrt{\vv{r} \cdot \vv{r}}$, and analogous the ''length'' of a function $f(\lambda)$ (i.e., its $l_2-$norm) is $|| \, f \, ||_2 = \sqrt{ \langle \, f \, | \, f \, \rangle}$. The angle $\beta$ between two vectors in ${\mathbb R}^3$ is given by $\cos(\beta) = \vv{r}_1 \cdot \vv{r}_2 \cdot ||\, \vv{r}_1\, ||_2^{-1} \cdot || \, \vv{r}_2\, ||_2^{-1}$, and in analogy the angle between two functions is given by $\cos(\beta) = \langle\, f_1\, | \, f_2\, \rangle \cdot ||\, f_1\, ||_2^{-1} \cdot || \, f_2\, ||_2^{-1}$. The familiar concept of Euclidian vectors thus allows for a good intuitive understanding of vector spaces of square-integrable functions. For introductions to functional analysis, the reader is referred to the rich supply of textbooks, e.g. the rather application oriented works by \citet{Milne1980} or \citet{Zeidler1995}. Using this mathematical formalism, in the following we formulate the problem of finding a passband $p(\lambda)$ as a problem of vector calculus.\par
The $N$ spectral energy distributions $s_i(\lambda)$ span a $M-$dimensional subspace of ${\mathcal L}^2(I)$, with $1 \le M \le N$. The extreme case $M=1$ holds if all $N$ sources have identical shapes in their SPDs, and only differ in brightness. The case $M=N$ holds if all $N$ SPDs are linearly independent, i.e. no $s_i(\lambda)$ can be expressed as a linear combination of the $N-1$ other SPDs. We assume $M$ to be known here, and postpone the question of how to determine this number in practice until Sec. \ref{sec:numberM}.\par
For the $M-$dimensional subspace, a set of $M$ basis functions, $\varphi_j(\lambda)\; j=1,\ldots,M$, exists, such that each $s_i(\lambda)$ can be expressed as a linear combination of the basis functions, and the basis functions are satisfying the orthonormality condition
\begin{equation}
\langle\, \varphi_k\, | \, \varphi_l\, \rangle = \delta_{kl}\; , \quad k,l = 1,\ldots,M \; , \label{eq:4}
\end{equation}
with $\delta_{kl}$ denoting the Kronecker delta. We discuss the question of how to find such an orthonormal basis for the subspace spanned by the $N$ SPDs in Sec. \ref{sec:basis}, and assume here that we already have available the set of $M$ basis functions, $\varphi_j(\lambda)$. Then we can express each of the $N$ SPDs by a linear combination of the basis functions, i.e.
\begin{equation}
s_i(\lambda) = \sum\limits_{j=1}^M\, a_{ij}\cdot \varphi_j(\lambda)\;, \; i=1,\ldots,N\quad . \label{eq:5}
\end{equation}
Combining the last equation with Eq. (\ref{eq:1}), we obtain
\begin{equation}
c_i = \sum\limits_{j=1}^M\, a_{ij}\cdot \int\limits_{\lambda_0}^{\lambda_1}\, p(\lambda)\, \varphi_j(\lambda)\, {\rm d}\lambda \equiv \sum\limits_{j=1}^M\, a_{ij}\, p_j\; , \; i=1,\ldots,N \quad ,
\end{equation}
with
\begin{equation}
p_j=\langle\, p\, | \, \varphi_j \, \rangle \quad \label{eq:coefDef} \quad .
\end{equation}
We can express the last equation more compactly as a matrix equation, writing the $N$ values $c_i$ as an $N \times 1$ vector $\bf c$, the values of all $a_{ij}$ as a $N \times M$ matrix $\bf A$, and the $M$ integral expressions $p_j$ as the $M \times 1$ vector $\bf p$,
\begin{equation}
{\bf c} = {\bf A}\, {\bf p} \quad . \label{eq:7}
\end{equation}
Now we observe that for an orthonormal basis, the elements of $\bf p$ are the coefficients of the passband $p(\lambda)$, developed in the basis functions $\varphi_j(\lambda), \; j=1,\ldots,M$. Thus, by knowing $\bf c$ and $\bf A$, we can solve Eq. (\ref{eq:7}) for the development of the passband in the basis functions that represent the SPDs available for the passband calibration. For this development of $p(\lambda)$ in the basis $\varphi_j(\lambda)$, we write $p_\parallel(\lambda)$,
\begin{equation}
p_\parallel(\lambda) = \sum\limits_{j=1}^M\, p_j \cdot \varphi_j(\lambda) \quad .
\end{equation}
The function $p_\parallel(\lambda)$ will in general not be identical with the passband $p(\lambda)$. To make $p_\parallel(\lambda)$ identical to $p(\lambda)$, it would be necessary that the subspace of ${\mathcal L}^2(I)$ spanned by the $N$ calibration spectra contains $p(\lambda)$ entirely. That is to say, for deriving the passband $p(\lambda)$ from the photometry of $N$ sources with known SPDs, it has to be possible that $p(\lambda)$ can be expressed as a linear combination of the $N$ SPDs $s_i(\lambda)\; , \; i=1,\ldots,N$. In general, this will not be the case. Since a photometric passband as a function of wavelength and astronomical SPDs are physically independent, one may not expect that among astronomical sources there exists any set of objects whose SPDs allow to represent some passband by linear combinations. As a consequence, the problem of reconstructing a passband from observations of astronomical sources with known SPDs is fundamentally limited to $p_\parallel(\lambda)$, which we call the {\it parallel component} in the remainder of this work, leaving the freedom of adding an {\it orthogonal component}, $p_\perp(\lambda)$. This orthogonal component is fulfilling orthogonality conditions with respect to all calibration sources,
\begin{equation}
\langle \, p_\perp \, | \,  s_i\, \rangle = 0 \; ,\quad i=1,\ldots,N \; . \label{eq:9}
\end{equation}
These orthogonality conditions imply that the orthogonal component of the passband is not contributing to the photometry of any calibration source. So we write for the passband
\begin{equation}
p(\lambda) = p_\parallel(\lambda) + p_\perp(\lambda)\quad , \label{eq:passbandDecomposition}
\end{equation}
where the parallel component is uniquely constrained by the available set of calibration sources and can be found by solving the system of linear Eqs. (\ref{eq:7}) for the vector $\bf p$, and where the orthogonal component is fully unconstrained by the calibration sources. The fact that $p_\perp(\lambda)$ is unconstrained by the calibration sources is a fundamental limitation in the determination of the passband $p$, introduced by the choice of the calibration sources. This component can only be guessed if the ''full'' passband $p$ is to be determined. This guessing can either be made explicit (as done in the remainder of this work) or implicit (if an initial passband guess is simply deformed in some way to obtain better agreement between observed and synthetic photometry).\par
An important difference between the parallel component $p_\parallel(\lambda)$ and the passband $p(\lambda)$ arises from the physical constraints on $p(\lambda)$. The passband has to be a non-negative function of wavelength. And as quantum efficiencies, mirror reflectivity, and filter transmissions tend to be rather smooth functions of wavelength, so is the resulting passband. ''Smooth'' in this work is understood in an intuitive way, meaning ''without too strong fluctuations on small scales'', and not meant in any strict mathematical sense. $p_\parallel(\lambda)$, and also $p_\perp(\lambda)$, however, do not have to fulfil these constraints individually, and in general they actually do not fulfil them. Smoothness and non-negativity apply only to the ''full'' passband $p(\lambda)$, i.e. the sum of the parallel and orthogonal component. With $p_\parallel(\lambda)$ determined by solving Eq. (\ref{eq:7}), the choice of $p_\perp(\lambda)$ is thus limited by the requirements of smoothness and non-negativity to the sum $p(\lambda)=p_\parallel(\lambda) + p_\perp(\lambda)$. We discuss the relevance of the orthogonal component, and methods for guessing, in more detail in the following section.

\section{Unconstrained passband component \label{sec:unconstrained}}
As the orthogonal component of the passband, $p_\perp(\lambda)$, is by construction not contributing to the photometry of the calibration sources, the choice made for $p_\perp(\lambda)$ is irrelevant for computing synthetic photometry of the calibration sources. And, as a consequence, it is also irrelevant for the synthetic photometry for all SPDs that can be expressed as a linear combination of the SPDs of the calibration sources. Such sources have SPDs that lie entirely in the subspace of ${\mathcal L}^2(I)$ that is spanned by the $N$ calibration sources. It may however be of interest to obtain synthetic photometry for sources with SPDs that have non-negligible components outside the subspace spanned by the SPDs of the calibration sources, i.e., sources whose SPDs cannot be well expressed as a linear combination of the $N$ SPDs of the calibration sources. The synthetic photometry of these sources depends on the choice for $p_\perp(\lambda)$. And as this component of the passband can only be guessed, the synthetic photometry becomes subject to systematic errors in the realistic case of an imperfect guess for $p_\perp(\lambda)$.\par
From these considerations, we may derive a criterion for a good choice of calibration sources in passband determination: The calibration sources are ideally chosen in such a way that the SPDs of the sources for which synthetic photometry is desirable can be expressed as a linear combination of the SPDs of the calibration sources. By doing so, systematic uncertainties resulting from ambiguities in the passband solution are minimised. In practice, this criterion can of course be met only imperfectly.\par
We now focus on the problem of obtaining an estimate for $p_\perp(\lambda)$. In principle, an orthonormal basis for the function space orthogonal to the set of calibration SPDs can be constructed easily by embedding the calibration SPDs in a higher-dimensional space. First, the empirical basis functions $\varphi_k(\lambda)$ are developed in a set of standard basis functions. A good choice are orthogonal systems of functions that are easily generated, such as Legendre polynomials. Functions which additionally provide simple analytic relations for the computation of their integrals make a particularly suitable choice, as they may simplify numerical computations. For this work, we found Hermite functions a useful way of changing from the empirical basis functions $\varphi_j(\lambda)$ derived from the observed SPDs to a mathematically more convenient representation. As Hermite functions are an orthogonal basis system on $\mathbb R$ instead of a finite interval, these functions require a continuation of the relevant functions with zero outside the considered interval. As a benefit, Hermite functions provide a set of basis functions which all converge to zero for arguments with sufficiently large absolute value, ideal for representing functions that, like passbands, are expected to smoothly approach zero for sufficiently large and small wavelengths. Hermite functions come with convenient recurrence relations for integral expressions, as discussed later. Additionally, the representation of the basis functions as a linear combination of functions that are in a strictly mathematical sense orthonormal simplifies computations and maintaining orthonormality by working in the coordinate space only.\par
After the $M$ basis functions $\varphi_j(\lambda)$, $j=1,\ldots,M$ are developed with sufficient precision in $L$ basis functions, usually with $L>>M$, and thus embedded in a higher-dimensional space, a singular value decomposition of the resulting $M\times L$ coefficient matrix allows to separate $M$ linear combinations of the $L$ basis functions representing (approximately) the basis functions $\varphi_j(\lambda)$, and $L-M$ linear combinations of the standard basis functions that provide a orthonormal basis for an orthogonal space. Any linear combination of the $L-M$ basis functions orthogonal to the basis functions $\varphi_j(\lambda)$ may be added to the parallel passband component without affecting the photometry of the calibration sources. As a linear combination of the $L$ standard basis functions allows for a non-negative and smooth linear combination, an orthogonal passband component satisfying the physical constraints in the passband when added to the parallel component, exists.\par
Finding such a linear combination in practice however is troublesome, as introducing smoothness conditions results in a very high dimensional non-linear optimisation problem. It is therefore desirable to restrict the variety of potential orthogonal passband components in exchange for finding only solutions that by construction satisfy the smoothness requirements. A way of doing so is introducing a smooth modification of an initial guess for the passband. Such an approach has basically been followed in previous works (Bessell 2000, Bessell \& Murphy 2012, Mann \& von Braun 2015, Ma{\'i}z Apell{\'a}niz 2017, among others). In the light of this work, however, the methods for modifying the initial guess for the passband are restricted by the requirement that it must be possible to modify the initial guess in such a way that the projection of the modified passband onto the subspace of ${\mathcal L}^2(I)$ spanned by the calibration sources can meet the solution for $p_\parallel(\lambda)$ obtained by solving Eq. (\ref{eq:7}). To achieve this, we assume an a-priori passband which is to be approximated, $p_{ini}(\lambda)$, and use a linear multiplicative model for modifying this initial passband, i.e. we write for the passband $p$ we are looking for
\begin{equation}
p(\lambda) = \left(\, \sum\limits_{k=0}^{K-1} \, \alpha_k\, \phi_k(\lambda)\, \right) \cdot p_{ini}(\lambda) \quad . \label{eq:15}
\end{equation}
If the functions $\phi_k(\lambda)$ are chosen to be smooth, and the initial passband $p_{ini}(\lambda)$ is smooth as well, then the resulting passband $p(\lambda)$ is smooth. Now we introduce the constraint that the function $p(\lambda)$ should have a certain parallel component $p_\parallel(\lambda)$ obtained, from solving Eq. (\ref{eq:7}), by using Eq. (\ref{eq:coefDef}). Putting Eq. (\ref{eq:15}) into Eq. (\ref{eq:coefDef}), we obtain
\begin{equation}
p_j = \langle\, \left(\, \sum\limits_{k=0}^{K-1} \, \alpha_k\, \phi_k(\lambda)\, \right) \cdot p_{ini} \, | \, \varphi_j\, \rangle \quad .
\end{equation}
From this equation, using all $j=1,\ldots,M$, we obtain a matrix equation
\begin{equation}
{\bf p} = {\bf M}\,{\boldsymbol \alpha} \quad , \label{eq:a17}
\end{equation}
with $\bf p$ from Eq. (\ref{eq:7}), the $K$ elements vector $\boldsymbol \alpha$ containing the coefficients $\alpha_k$ of the multiplicative modification model, and an $M\times K$ matrix $\bf M$, with
\begin{equation}
{\bf M}_{n,m} = \langle\, \phi_m\, p_{ini} \, | \, \varphi_n\, \rangle \quad .
\end{equation}
Thus, given some initial passband $p_{ini}(\lambda)$ and choosing some linear modification model $\phi_k$, $k=0,\ldots,K-1$, from computing the matrix $\bf M$ we can immediately see if a solution $p(\lambda)$ exists that satisfies our constraints on $p_\parallel(\lambda)$, and if this modification is unique. The coefficients for this modified passband, $\boldsymbol \alpha$, are obtained from solving Eq. (\ref{eq:a17}).\par
We will consider two useful aspects of finding a passband $p(\lambda)$ from modifying an initial passband $p_{ini}(\lambda)$ under the constraint on the parallel component $p_\parallel$. First, we may choose $K>M$, i.e. using more free parameters to modify the initial passband than we have constraints by the parallel component. In this case, the linear system of Eq. (\ref{eq:a17}) becomes underdetermined. If we determine a basis of the $K-M$-dimensional null space of the matrix $\bf M$, we can add any linear combination of this basis to the coefficient vector $\boldsymbol \alpha$, and obtain a modified initial passband that satisfies the constraint on $p_\parallel(\lambda)$. This way, we easily introduce some free tuning parameters that can be used to make the solution for $p(\lambda)$ non-negative, should the exact solution not be so, or to tune the passband solution $p(\lambda)$ more to our a-priori knowledge or prejudice on how the passband should look like.\par
Second, the coefficients for the parallel component, $\bf p$ are affected by typically considerable errors. Fixing $\bf p$ exactly to the least squares solution of Eq. (\ref{eq:7}) however may not be optimal. The solution of Eq. (\ref{eq:a17}) has to balance out the fine structure in the parallel component of the passband delicately, and a slight change in $\bf p$, within the uncertainties resulting from Eq. (\ref{eq:7}), may result in a solution of $p(\lambda)$ that is much closer to the desired initial passband, with but a minute decrease in the goodness of fit in Eq. (\ref{eq:7}). To take this possibility into account, we may compute the formal variance-covariance matrix on $\bf p$ from Eq. (\ref{eq:7}), and then draw multivariate normally distributed random vectors $\bf p_r$ from that variance-covariance matrix. For each random sample $\bf p_r$, one solves Eq. (\ref{eq:a17}) for the passband $p(\lambda)$. From a number of random samples, one can choose the solution that is non-negative and closest (in an $l_2$-sense) to $p_{ini}(\lambda)$. This random sampling approach might be brute, but as the numerical effort involved is small, the range of parallel passband components within the uncertainty boundaries of the exact solution $\bf p$ is easily well sampled.\par
In this work, we use a polynomial modification model, i.e. $\phi_k = \lambda^k$, as a convenient but eventually arbitrary choice. As we use a development of the basis functions $\varphi_j$, as well as of the initial passband $p_{ini}(\lambda)$, in Hermite functions, this polynomial modification model allows us to use the orthonormality and the relation
\begin{equation}
x\, \bar{\varphi}_n(x) = \sqrt{\frac{n}{2}} \, \bar{\varphi}_{n-1}(x) + \sqrt{\frac{n+1}{2}}\, \bar{\varphi}_{n+1}(x)
\end{equation}
for Hermite functions $\bar{\varphi}_n$ iteratively to compute the integrals $\langle\, \phi_m\, p_{ini} \, | \, \varphi_n\, \rangle$ analytically.\par
We use both the random sampling approach and the null space method, to adjust the passband solutions to the initial passbands and obtain what we consider the preferable solution for the HIPPARCOS, Tycho, and {\it Gaia} DR1 passbands, as discussed in more detail in Sec. \ref{sec:passbands}.

\section{Basis functions \label{sec:basis}}

Until now, we have assumed the basis functions $\varphi_j(\lambda)$, $j=1,\cdots,M$, and the number of basis functions, $M$, to be known. In the following, we discuss the construction of the basis functions, and a practical choice for the number of basis functions.

\subsection{Construction of basis functions}

   \begin{figure}
   \centering
   \includegraphics[width=0.5\textwidth]{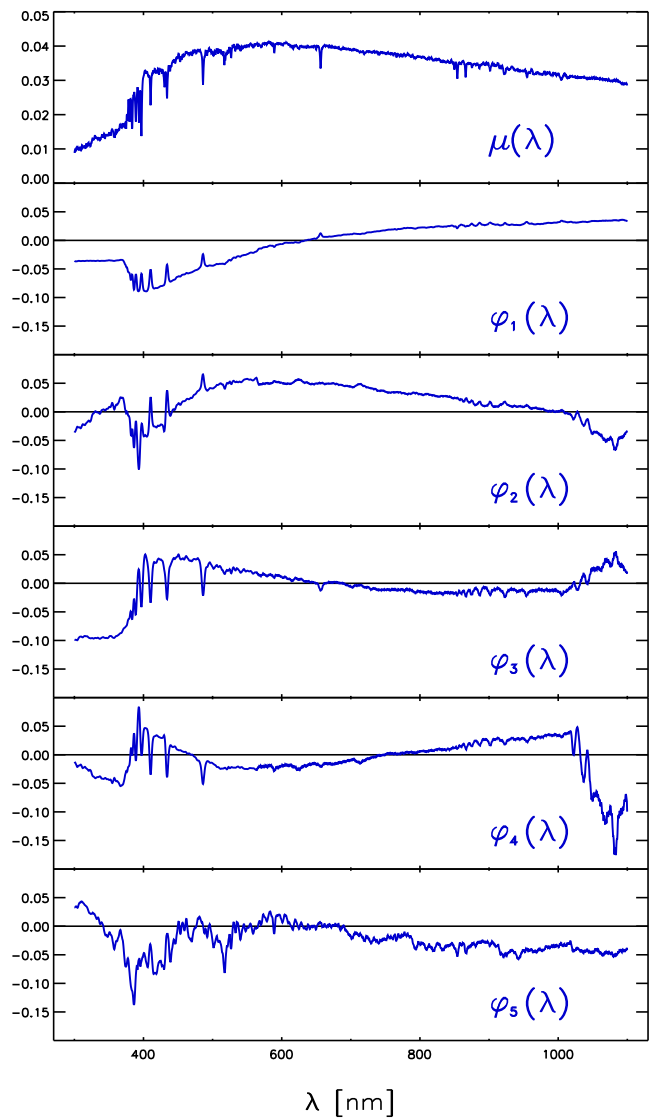}
   \caption{The mean function $\mu(\lambda)$ and the first five eigenfunctions $\varphi_k(\lambda)$, $k=1,\ldots,5$, for the 124 calibration sources we use here for the {\it Gaia} passband.}
              \label{Fig:1}
    \end{figure}

Given a set of $N$ SPDs defined on the wavelength interval $I$, an orthonormal basis for these SPDs can be constructed using functional principal component analysis \citep{Ramsay2006}. In this approach, each SPD is represented by a truncated Karhunen-Lo{\`e}ve (KL) representation,
\begin{equation}
s_i(\lambda) = \mu(\lambda) + \sum\limits_{j=1}^M\, \xi_{i,j} \cdot \varphi_j(\lambda) \quad . \label{eq:kl}
\end{equation}
Here, $\mu(\lambda)$ is the mean function derived from all calibration sources. The functions $\varphi_j(\lambda)$ are the eigenfunctions of the covariance operator, and as such they fulfil orthonormality. The coefficients $\xi_{i,j}$, usually called the {\it scores} of the function $s_i(\lambda)$, provide the development of the calibration source $i$ in the basis functions $\varphi_j$, $j=1,\ldots ,M$.\par
In practice, the SPDs of the calibration sources are usually available in tabulated form, specifying the fluxes at a number of discrete wavelength points. We may consider the possibility of a number of complications that happen to occur in practice. These complications are:
\begin{itemize}
\item The wavelength sampling not being identical for all calibration sources
\item The wavelength sampling changing within a spectrum
\item The wavelength sampling being non-linear
\item Missing values or even gaps within a tabulated SPD
\item The sampled flux values of the SPD are affected by random noise
\end{itemize}
To account for such circumstances when estimating $\mu(\lambda)$ and the $\varphi_j(\lambda)$, we follow the approach developed by \citet{Yao2005}.\par
For applying the methods lined out in previous sections in the determination of passbands for HIPPARCOS, Tycho, and {\it Gaia}, we require a set of calibration sources with known SPDs over the whole wavelength range where the passbands have to be assumed non-zero. The set of calibration sources should be homogeneous in wavelength resolution. We use spectra from the Next Generation Spectral Library \citep{HeapLindler}, NGSL in the following, Version-2, for this purpose. The spectra were taken with the STIS spectrograph onboard the Hubble Space Telescope, and they cover the wavelength range from the ultraviolet to about 1030 nm. On the wavelength interval of interest to this work, from 290 nm upwards, the spectra were obtained with two different gratings, meeting at about 565~nm. The spectral resolving power of these spectra ranges from about 530 to 1170 \citep{HeapLindler}. The influence of the wavelength resolution onto the results of this work is discussed in more detail in Sec. \ref{sec:wavelengthResolution}.\par
All spectra in the NGSL with HIPPARCOS, Tycho, or {\it Gaia} DR1 measurements were selected. Stars that were indicated as variable either in the Simbad data base or in the HIPPARCOS catalogue were excluded from the use as calibration sources. Furthermore, a few sources which were clearly outliers in the synthetic photometry were also excluded. As an additional criterion, sources fainter than 10.5 mag in $H_p$ and $V_T$, and fainter than 11.5 mag in $B_T$ were excluded from the determination of the HIPPARCOS and Tycho passbands, as we found indications of systematic trends in the HIPPARCOS and Tycho-2 photometry for larger magnitudes. Finally, the sets of calibration sources include $N=$ 210 stars for Tycho B and V passbands, $N=$ 206 stars for the HIPPARCOS passband, and $N=$ 124 stars for {\it Gaia} DR1 passband. For the calibration of the {\it Gaia} data, a particular, optimised set of calibration sources has been developed by \citet{Elena2012}, and used internally by the Gaia Data Processing and Analysis Consortium. As the SEDs of this calibration set are not publicly available by now, we restrict the calibration of the $G$ passband to the NGSL spectra.\par
Another suitable set of calibration spectra are the CALSPEC sources \citep{Bohlin2017}. The number of CALSPEC spectra of sources sufficiently bright for a calibration of $H_p$ and the Tycho passbands is however small compared to the number of NGSL spectra available, and these CALSPEC sources cover essentially the same spectral types as the NGSL data set does. A larger number of CALSPEC  sources could be added for the calibration of the $G$ passband. The fainter CALSPEC sources include also two M-type stars, a spectral class absent in the NGSL set of calibration sources used in this work, and which could provide an extension of the subspace spanned by the calibration sources. However, as is discussed in Sec.~\ref{sec:Gaia}, we found indications for a systematic trend in the $G$ magnitude, which makes an extension of the magnitude range of the calibration sources for the $G$ passband problematic. In particular, including additional spectral types only in a different magnitude range might result in systematic distortions of the passband, rather than improving it. We therefore rather restrict this work to a homogeneous set of calibration sources, using only NGSL spectra. The sets of calibration sources used in this work contain stars of spectral types O, B, A, F, G, and K. A list of the stars used as calibration sources, and their properties, is provided in an online table.\par
The wavelength range covered by the NGSL spectra, reaching up to 1030~nm, includes the transmission ranges of the $H_p$, $B_T$, and $V_T$ passbands. Only the $G$ passband is expected to be non-zero at wavelength larger than 1030~nm. We therefore expand the wavelength range up to 1100~nm by fitting the NGSL spectra with a linear combination of BaSeL spectra \citep{BaSeL} similar to the NGSL spectrum, and using the obtained fit as an extrapolation from 1030~nm up to 1100~nm. This procedure is clearly a crude estimate for the continuation of the spectra. But as the $G$ passband is already very close to zero at the wavelengths for which the spectra have to be extrapolated, such a simplistic extension of the spectra will do.\par
Fig.~\ref{Fig:1} shows the mean function $\mu$ and the first five eigenfunctions $\varphi_k$, $k=1,\ldots,5$, for the case of the calibration set for {\it Gaia} DR1, computed over the wavelength interval $I=[300,1100]$ nm, as an example for the procedure lined out here. It should be noted that all the shown functions have a purely descriptive character for the set of SPDs used, and must not be interpreted in physical terms. Typical features of the eigendecomposition can be noted, such that the lower eigenfunctions tend to be smoother, representing more general trends in the SPDs, while with increasing number, the eigenfunctions become more complex in structure, adding more details in wavelength to the linear combination of eigenfunctions. As we are interested only in representing the SPDs of the calibration set by a set of orthonormal functions, we orthonormalise the mean function with respect to the eigenfunctions, and add the result to the set of basis functions. The wavelength interval may be adjusted to the expected range of sensitivity for other passbands, resulting in bases covering different, narrower intervals in wavelength for $H_p$, $B_T$, and $V_T$.

\subsection{Fixing the number of basis functions \label{sec:numberM}}

Until now, we have considered the number of basis functions, $M$, to be known. In practice, this number is not clearly defined, as any set of $N$ basis functions will not be described {\it exactly} by a set of basis functions less than $N$ (the noise alone prevents this). When representing a set of SPDs with an empirical basis, we observe rather a better and better approximation to all SPDs in the set with increasing number of basis functions used, but the improvements in approximations become smaller and smaller. At some point, one may consider the use of further basis functions irrelevant, as they will start to approximate the noise in the empirical SPDs. Furthermore, even if the improvement in the representation of the SPDs is still significant, considering the uncertainties on the SPDs, it may however already be insignificant for calculating photometry, considering the uncertainties in photometry. This is caused by the low sensitivity of a weighted integral over SPDs with respect to small-scale variations in the SPD.\par
A simple approach for determining the number of basis functions would therefore be to add basis functions until the level of residuals is statistically in good agreement with the level of residuals expected from the uncertainties on the data used. The good agreement can be judged by a statistics test, provided that a reliable error model is available. As discussed in detail in the following section, for this work, however, we have to expect a significant impact of the uncertainties on the SPDs onto the residuals in passband determination, without having a quantitatively reliable error model for the SPDs available. We therefore have to estimate the value for $M$ in this work, by adding more basis functions and considering the increase in the formal error on the coefficient for the added basis function, the overall decrease in the goodness of fit parameter (i.e., $\chi^2$ in this work), and the distribution of residuals. If the formal errors on the coefficients start to strongly increase, the improvement in the overall goodness of fit becomes very small, and the residuals begin to change only for one, or very few, stars in the calibration set, we assume that the additional basis function is not well constrained anymore. In this case, further improvements in the goodness of fit are achieved only by adjusting to features peculiar to individual stars, and therefore probably spurious. By doing so, we finally consider the use of $M=4$ for $H_p$ and $B_T$, and $M=3$ for $V_T$ and $G$, to be reasonable choices for the number of basis functions.

\section{Effects of uncertainties \label{sec:uncertainties}}

When solving Eq. (\ref{eq:7}) for the parallel component of the passband, we have to consider different effects of uncertainties. First, we have errors on the observed photometry, i.e. on the vector $\bf c$. When working with photometric observations which reach very low errors, we may also have to consider the effect of uncertainties on the SPDs used for determining the passband, i.e. errors on the matrix $\bf A$ in Eq. (\ref{eq:7}). The uncertainties in the SPDs are assumed to consist of two contributions, an error in the absolute flux level, and an error in the shape of the SPDs. In the following, we discuss the different kinds of uncertainties in more detail.

   \begin{figure}
   \centering
   \includegraphics[width=0.5\textwidth]{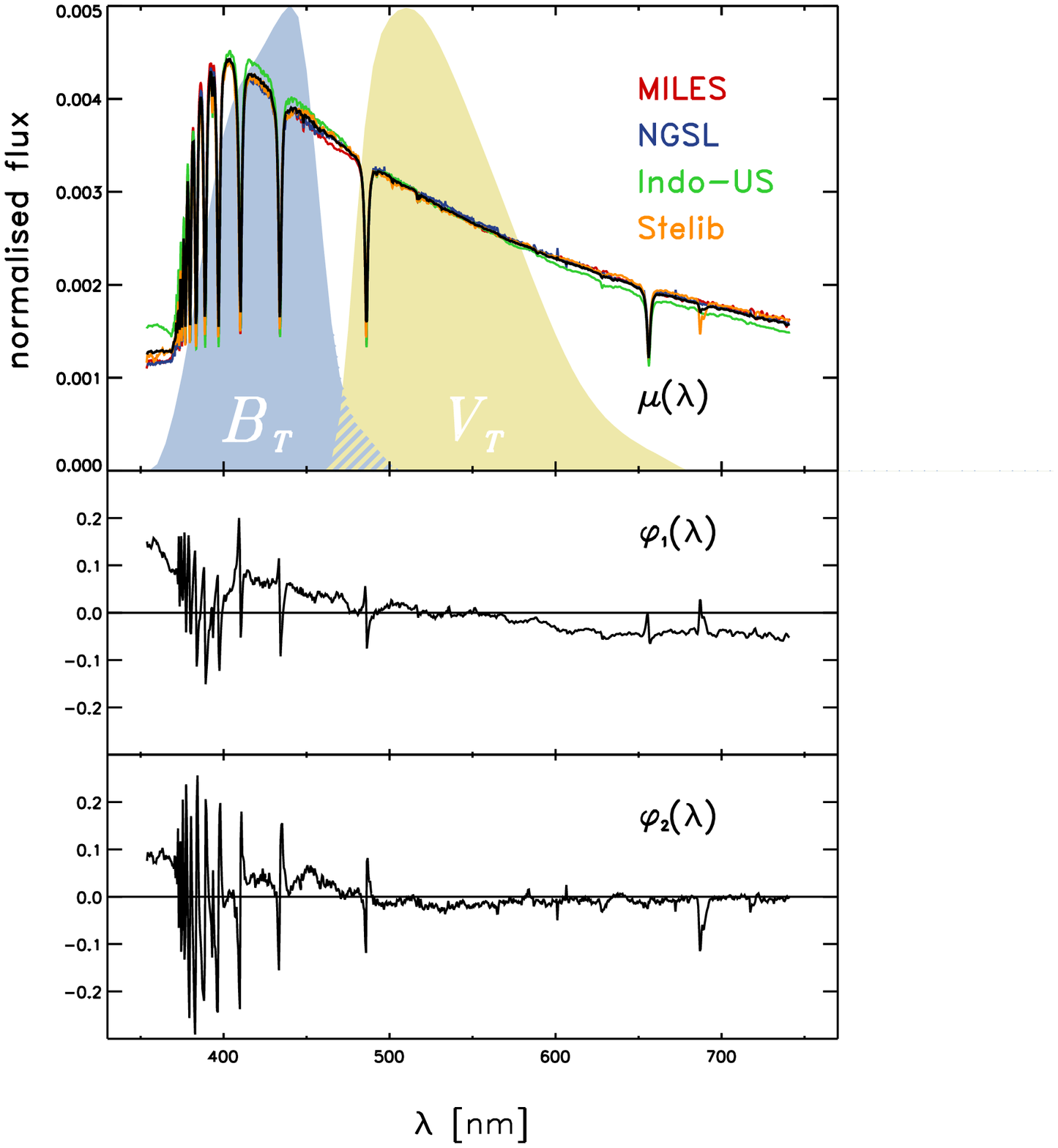}
   \caption{Example for the estimation of the uncertainty in SPD shape for star HD 109995. Top panel: $l_1$-normalised SPDs from four different libraries, plus the derived mean SPD, $\mu(\lambda)$. The passbands for $B_T$ and $V_T$ according to \citet{ESA1997} are shown schematically for comparison. Central and bottom panels: The first and second eigenfunctions, $\varphi_1(\lambda)$ and $\varphi_2(\lambda)$, derived from the four spectra of HD 109995.}
              \label{Fig:2}
    \end{figure}

\subsection{Error in observed photometry}

For the errors on the observed photometry, one may assume statistical independence between different sources and passbands. Quantitative estimates for the errors are available for all photometric measurements used in this work. These photometric measurements cover a wide range of magnitudes, resulting in very different signal-to-noise ratios. Furthermore, for {\it Gaia}, the number of observations for each source also varies strongly, again resulting in very different signal-to-noise ratios. A weighting of the measurements with the error would therefore result in a few bright and/or frequently observed stars dominating the passband determination, thus reducing the coverage of spectral types. As the random error on the photometric measurements are, as discussed in the following sub-sections, in many cases not the dominant source of uncertainty in the passband determination, a weighting with the photometric error would be an unsuitable strategy. Instead, we neglect this error and apply the same relative weight to each photometric data point.

\subsection{Error in flux level}

The error on the absolute flux level of a spectrum is typically in the order of percent. In case of the NGSL, this uncertainty was estimated to about 2 to 3\% by \citet{HeapLindler}. So this uncertainty is dominant when considering photometric measurements with milli-magnitude precision. The uncertainty in the flux level affects all synthetic photometry derived from a given SPDs in the same way, i.e. it results in the same relative error in the synthetic photometry in all passbands, with a correlation of $\rm +1$ between passbands. As a consequence, this uncertainty cancels out in first order when computing the ratio of the synthetic photometry of the same SPD in two different passbands. We write $f = c_1/c_2$ for this ratio, and obtain for the error in $f$
\begin{equation}
\sigma_f^2 \approx {\bf J} \, \Sigma_c \, {\bf J}^{\mathsf T} = 0 \quad . \label{eq:absError}
\end{equation}
Here, $\bf J$ denotes the Jacobian of the function $f$, and $\Sigma_c$ is the variance-covariance matrix for the photometry $c_1$ and $c_2$, in two passbands, describing the error in the absolute flux level. With the same relative error $\sigma$ on $c_1$ and $c_2$, and a correlation coefficient $\rho = 1$, this matrix is given by
\begin{equation}
\Sigma_c = \sigma^2\,
\begin{pmatrix}
c_1^2 & c_1\, c_2 \\
 c_1\, c_2 & \, c_2^2 \\
\end{pmatrix}
\quad .
\end{equation}
The validity of Eq. (\ref{eq:absError}) can be seen by simply putting in the expressions for $f$ and $\Sigma_c$. As far as the error in the absolute flux level is concerned, we are thus able to predict the ratio between the synthetic photometry in different passbands with higher relative accuracy than the synthetic photometry itself. As demonstrated in Sec. \ref{sec:passbands}, we indeed obtain relative residuals on a level even sightly less than 2\% for the most accurate photometry, with a strong positive correlation between passbands, and significantly lower relative error on the ratios of fluxes in different passbands.

\subsection{Theoretical treatment of errors in the shape of SPDs}

We now consider possible errors in the shape of SPDs. Uncertainties in wavelength calibration, response function, or, for ground-based observations, in the atmospheric extinction correction may introduce systematic differences between the SPD as derived from spectroscopic observations, and the true SPD. When computing integral expressions over the SPD, such as synthetic photometry, such systematic differences can amplify, depending on the scales of the deviation and the weighting function, i.e. the passband, used in the integration. A convenient way for quantifying such uncertainties in shape of the SPD, and for propagating the uncertainties into inner products with the SPD, is again the use of functional principal component analysis. The approach is similar to the approach we were using for the construction of the basis functions in Sec. \ref{sec:basis}, but follows a different intention here. Assuming we have a number of measured SPDs of the same source available (and the source is stable in time), we can interpret the different observed SPDs as random realisations of the true SPD. For this random process we can again use the Kahunen-Lo{\`e}ve representation given by Eq. (\ref{eq:kl}). Different from Sec. \ref{sec:basis}, where we were interested only in the eigenfunctions $\varphi_j(\lambda)$ as a convenient way for representing a set of functions, we now make use of the fact that for a random process, the scores $\xi_j$ in Eq. (\ref{eq:kl}) are uncorrelated random variables with a variance given by the eigenvalues $\nu_k$ corresponding to the $k$-th eigenfunction $\varphi_k(\lambda)$. We thus derive the mean function $\mu(\lambda)$ from all observations of the SPD of the same source as the best estimate for the true SPD, and quantify the uncertainty on the mean function by a finite series of eigenfunctions $\varphi_j(\lambda)$, $j=1,\ldots,J$, where the contribution of each eigenfunction is a random variable with variance $\nu_j$. From this, we can compute the error on any inner product with the SPD. To do so, it is necessary to assume some probability density distribution for the random variables determining the contribution from each eigenfunction, as the Karhunen-Lo{\`e}ve theorem constraints only the variances of these random variables, and guarantees they are uncorrelated. It makes however no further statement about their probability distribution. In lack of any better information, we make the usual assumption of normal distributions here.\par
For an inner product $y$ of some function $p(\lambda)$ with the SPD $s(\lambda)$, we thus obtain
\begin{equation}
y  =  \langle\, p \, | \, s\, \rangle = \langle\, p \, | \, \mu\, \rangle + \sum\limits_{j=1}^J\, \xi_j \cdot \langle\, p\, | \, \varphi_j\, \rangle \quad ,
\end{equation}
with $\xi_j$ being independently normally distributed random variables with variances $\nu_j$. The variance of $y$ is thus given by
\begin{equation}
\sigma_y^2 = \sum\limits_{j=1}^J\, \nu_j \cdot \langle\, p\, | \, \varphi_j\, \rangle^2 \quad .
\end{equation}
We now consider the case that we derive a number of $n$ different inner products with the same SPD, $y_i = \langle\, p_i \, | \, s\, \rangle$, $i=1,\ldots,n$. We write the $y_i$ as an $n\times 1$ matrix $\bf y$. The inner products with the eigenfunctions, $\langle\, p_i\, | \, \varphi_j\, \rangle$ we write as an $n \times J$ matrix $\bf P$. We then obtain for the variance-covariance matrix of $\bf y$, named $\Sigma_{\bf y}$, the expression
\begin{equation}
\Sigma_{\bf y} = {\bf P} \,\Sigma_\nu \,{\bf P}^{\mathsf T} \quad , \label{eq:17}
\end{equation}
where $\Sigma_\nu := {\rm diag}(\nu_j)$, the diagonal matrix containing the $J$ eigenvalues. Thus, making use of the Karhunen-Lo{\`e}ve formalism, we can derive the variances of the inner product, as well as the correlations with other inner products. If the function $p(\lambda)$ represents a passband, this corresponds to estimating the variance of the integrated photometry, as well as the correlation between the errors of the integrated photometry in different passbands.

\subsection{Estimates for the error on the shape of SPDs}

To illustrate the results in practice, we require a number of independent observations of SEDs of the same sources. To find these, we select from the set of NGSL calibration sources the ones which are also included in other spectral libraries. The libraries considered here are the MILES library \citep{MILES}, the Stelib library \citep{Stelib}, the X-shooter library \citep{XSL}, and the Indo-US library \citep{Valdes2004}. We selected 13 stars which are included in four out of these five spectral libraries, and use the four independent determinations of the SPDs of each of these stars as the basis for determining the uncertainty in the shape of the SPDs. Four measurements of the SPD per source is a rather small data base for reliably estimating the uncertainty in the shape. However, for illustration of the principle and an estimate of the order of magnitude of the effects of uncertainties in shape, it will do.\par
As the five spectral libraries used provide spectra with different spectral resolution, we first adapt the wavelength resolution to the library with the lowest one, which is NGSL. To do so, SPDs from each library are convolved with a Gaussian and the shape of lines in a SPD is compared with the shape of the same line in the NGSL library. The width of the Gaussian is adapted until a visually good agreement in line width is achieved. This process is done for many lines at different wavelengths, and the resulting Gaussian widths as a function of wavelength have been fitted with polynomials (separately for each wavelength range in the NGSL spectra covered by a particular grating). Finally, all SPDs are convolved with a Gaussian with variable widths, according to the fitting done for each library. The result is a fairly good agreement of the spectral resolution of all spectra from all libraries over the entire wavelength range common to all five libraries.\par
For the MILES spectral library, no absolute flux calibration is available. Furthermore, the precision of the absolute flux calibration for the different spectral libraries is different. We therefore normalise all spectra to unit flux over the common wavelength range before applying the KL decomposition. Using this normalisation with respect to the $l_1$ norm it is made sure that the spectra only differ in the shape of the SPDs.\par
Fig.~\ref{Fig:2} shows the result for one example source, HD 109995. The top panel shows the individual normalised SPDs, as obtained from NGSL, MILES, Indo-US, and Stelib spectral libraries, together with the derived mean function $\mu(\lambda)$, over the wavelength interval common to all four libraries. The central and bottom panels show the first and second eigenfunctions, $\varphi_1(\lambda)$ and $\varphi_2(\lambda)$, respectively, derived from the four measurements of the SPDs. Several features of this decomposition may be noted:\par
Both the mean function and the eigenfunctions contain considerable small scale fluctuations by noise. This is the result of the low level of smoothing that has been applied before the KL decomposition. Stronger smoothing (i.e. larger bandwidths for the data and covariance estimate local linear smoothers in the formalism by \citet{Yao2005}) would have resulted in smoother curves. However, the spectra contain absorption lines with rather narrow peaks, the representation of which would have become worse with larger smoothing. Here, one is in conflict between suppression of noise and precision of result, as the data contains true features on the same wavelength scale as the noise. A rather conservative compromise was chosen in this work, which means that the fine structures in the resulting mean function and eigenfunctions should not be interpreted. For $\varphi_1(\lambda)$, providing the larger contribution to the uncertainty in the shape of the SPD, we thus find a rather simple general trend at wavelengths larger than about 410~nm. This trend is superimposed by local zigzag patterns which coincide with the position of absorption lines in the spectra. These patterns correct for mismatches in the position of the spectral lines between the four input SPDs, i.e. the uncertainty in wavelength calibration. At wavelengths shorter than about 410~nm, the Balmer lines become so close that the eigenfunctions take a very complex appearance resulting from the superposition of the zigzag features. The general trend is also broken, indicating more complex differences between the input spectra at short wavelengths. This reflects the difficulties of obtaining the good spectrophotometric calibration at the very blue end of the visual spectral range for ground-based observations. Finally, we recall that the eigenfunctions are normalised, and magnitude of uncertainty in the shape of the SPD and the contributions of each of the two eigenfunctions cannot be inferred from Fig.~\ref{Fig:2}, but are determined by the corresponding eigenvalues. These eigenvalues are $\nu_1 = 4.12\times 10^{-6}$ and $\nu_2 = 3.23\times 10^{-6}$ in the example shown in Fig.~\ref{Fig:2}.\par
The wavelength range common to all five spectral libraries used in this work contains the passbands $B_T$ and $V_T$ completely. These passbands, according to \citet{ESA1997}, are shown schematically in Fig.~\ref{Fig:2} for illustration. Thus, we can use Eq. (\ref{eq:17}) to estimate the relative errors (i.e. the square root of the variance, divided by the photon flux) and the correlation coefficient for the $B_T$ and $V_T$ passbands for the 13 test stars in the data set. The results, assuming the passbands by \citet{ESA1997}, are presented in Table~\ref{tab:1}. These numbers have to be interpreted with care, as the data set of four measurements of each SPD is rather small, and some spectra show significant deviations from the remaining three spectra for the same source, i.e. some kind of outlier behaviour. The mean and median error of the $B_T$ and $V_T$ passbands derived from the small set of test sources is presented in Table \ref{tab:1}, too. Both are on a percent level. The estimated uncertainty for the $B_T$ passband is rather high, though, and not in agreement with the lower level of residuals found in the $B_T$ passband determination described in Sec. \ref{sec:BT}. A reason for this might be the different qualities of the spectrophotometric measurements that enter into the error estimation. The NGSL spectra, exclusively used for the passband determination, are taken from space without the disturbing effects of the atmosphere, and therefore may have a higher precision than those estimated from a set of SPD measurements containing ground-based observations.\par
The variations observed between the SPDs of the same star are mostly a simple trend with wavelength (cf. the first eigenfunction, $\varphi_1(\lambda)$, shown in the example in Fig.~\ref{Fig:2}), and therefore affects the $V_T$ passbands, located around the centre of the wavelength interval considered, less than the $B_T$ passband. The uncertainties found for this passband, and listed in Table~\ref{tab:1}, are consequently lower, and in better agreement with the residuals found in the determination of the $V_T$ passband. Taking the lower errors of the $V_T$ passband, we still obtain errors introduced by the uncertainty in the shape of the SPD which are not negligible compared to the uncertainties in the observed photometry. These errors also show a strong correlation between different passbands for the same SPD. We therefore have to assume that not only the uncertainty in the absolute flux level of the SPDs, but also the uncertainty in the shape of the SPDs, is relevant in passband determinations from high-precission photometric measurements.

\subsection{Mathematical treatment of errors}

With results from the discussion of the errors on the absolute flux level and errors on the shape of the SPD, we are in the position that the solution for the passband, by solving Eq. (\ref{eq:7}), is strongly affected by noise on the matrix elements, resulting from uncertainties in the shape of the SPDs used in the calibration process, and not only on noise on the vector elements on the right-hand side, as typically assumed in the solution of linear equations. Suitable formalisms and algorithms for handling such situations are available, e.g. the element-wise weighted least squared (EWLS) approach by \citet{Markovsky2006}. The stringent use of such approaches however requires a reliable quantitative knowledge of the KL decomposition of the SPDs for all calibration stars used. This knowledge is not available, as there are rarely many independent measurements of the SED of the same source published, which could serve as a basis for reliably estimating the uncertainty in the shape of the SPD. We are therefore not able to provide a stringent treatment of the uncertainties in the computations of the passbands. As a simple test, we selected the HIPPARCOS data set, having the lowest errors on the observed photometry, and computed the passband solution from Eq. (\ref{eq:7}) using the EWLS formalism, and compared the result with the solution obtained with the ordinary least squares approach. For the error on the matrix elements, we were assuming a relative error of 3\%, maximally correlated within each row of the matrix $\bf A$ in Eq. (\ref{eq:7}), and uncorrelated between columns of $\bf A$, and $\bf p$. The result agreed well within the formal errors of the ordinary least squares result, the dominant effect of the errors on $\bf A$ being a shift in the residuals of about 1 milli-magnitude.\par
We may thus use the standard least squares solution to Eq. (\ref{eq:7}) as a good approximation. We should however keep in mind that the level of residuals in reproducing the observed photometric data will not be determined only by the uncertainty on the photometric data, but also by the uncertainty in the shape of the SPDs used in the calibration process. When different passbands are determined from the SPDs of the same calibration stars, we should furthermore expect correlations between the residuals for the individual passbands.

\begin{table}
\begin{center}
\renewcommand\arraystretch{1.2}
\caption{\label{tab:1}The relative errors on the photon counting rates (in \permil) in the nominal $B_T$ and $V_T$ passbands, and the corresponding correlation coefficient $\rho$, for the 13 stars with four independent determinations of the SPDs per star. The spectral types are taken from the HIPPARCOS catalogue.}
\begin{tabular}{| r | r | c | r | r | r |} \hline
\multicolumn{2}{|c|}{Star ID} & Spectral & \multicolumn{2}{c|}{Relative error} & \multirow{2}{*}{$\rho$} \\
\multicolumn{1}{|c}{HD} & \multicolumn{1}{c|}{HIP} & type &  \multicolumn{1}{c}{$B_T$} & $V_T$ & \\ \hline
2857 & 2515 & A2 & 7.9 &16.7 & $-$0.25 \\ % 7
25329 & 18915 & K1V & 100.1 & 4.3 & 0.92 \\ % 51
28978 & 21295 & A2Vs & 76.8 & 25.3 & $-$0.92 \\ % 8
37828 & 26740 & K0 & 24.2 & 16.3 & 0.999 \\ % 55
45282 & 30668 & G0 & 71.0 & 17.7 & 0.98 \\ % 33
58551 & 36152 & F6V & 26.6 & 10.2 & $-$0.64 \\ % 220
94028 & 53070 & F4V & 24.7 & 25.1 & 0.72 \\ % 254
105546 & 59239 & G2IIIm & 172.8 & 43.0 & 0.85 \\ % 263
106038 & 59490 & F6V-VI & 179.3 & 28.0 & 0.94 \\ % 143
109995 & 61696 & A0p & 25.9 & 7.2 & $-$0.50 \\ % 268
148513 & 80693 & K4IIIp & 43.0 & 16.8 & 0.89 \\ % 302
175305 & 92167 & G5III & 340.5 & 33.2 & 0.37 \\ % 324
175640 & 92963 & B9III & 34.5 & 15.0 & $-$0.59 \\ \hline % 5
\multicolumn{2}{|c|}{mean} & &  86.7 & 19.9 & \\
\multicolumn{2}{|c|}{median} & & 43.0 & 16.8 & \\ \hline

\end{tabular}
\end{center}
\end{table}

\section{Passbands for HIPPARCOS, Tycho, and Gaia DR1 \label{sec:passbands}}

Equipped with the formalism lined out so far, we now determine the passbands for $H_p$, $B_T$, $V_T$, and $G$. In all cases, we determine the parallel component of the passband from solving Eq. (\ref{eq:7}), and then using the described techniques to estimate the orthogonal component of the passband such that the solution approximates an initial passband guess.\par
The passbands derived in this work and discussed in this section are available as an online table, containing the parallel and orthogonal components, respectively, as a function of wavelength.

\subsection{HIPPARCOS passband \label{sec:hipparcos}}

\begin{figure}
   \centering
   \includegraphics[width=0.5\textwidth]{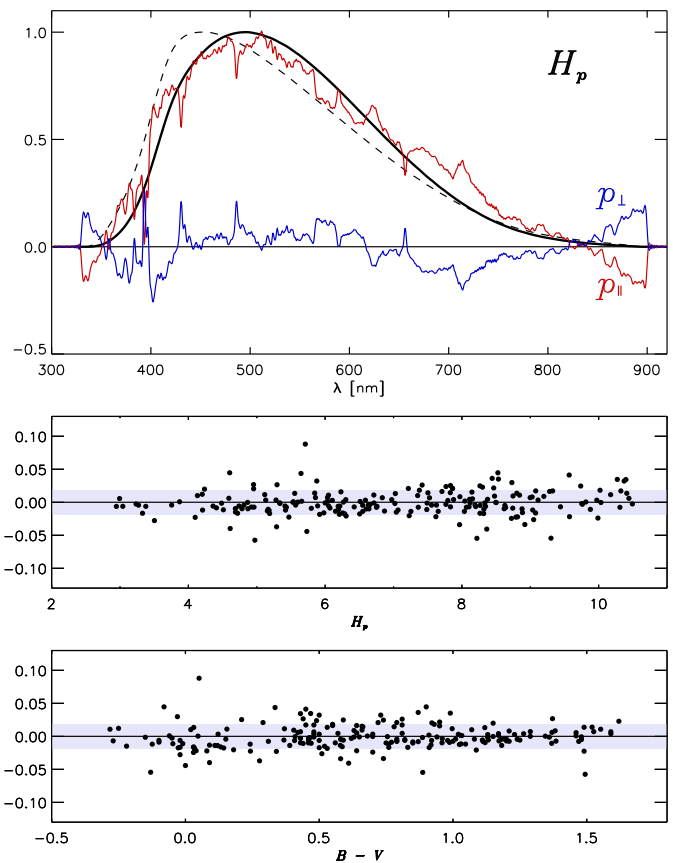}
   \caption{Passband solution for $H_p$. Upper panel: The parallel and estimated orthogonal components, $p_\parallel$ and $p_\perp$, shown as red and blue curve, respectively. The thick solid line is the sum $p$. The dashed line shows the initial passband $p_{ini}$, which was used for estimating the orthogonal component. The middle panel shows the relative flux residuals of the calibration stars versus the HIPPARCOS magnitude, the lower panel versus the Johnson $B-V$ colour. The highlighted areas indicate a range of one standard variations.
              \label{Fig:3}}
    \end{figure}

The HIPPARCOS passband has been derived from 206 NGSL spectra. After deriving $p_\parallel$ from the solution of Eq. (\ref{eq:7}), the random sampling approach has been used to find a passband solution close to the passband as provided by \citet{ESA1997} as the initial guess, $p_{ini}$. A polynomial of degree 3 for the modification of $p_{ini}$ was employed. The result is shown in Fig.~\ref{Fig:3}. The upper panel in this figure shows the parallel and estimated orthogonal component, together with the sum of both components. The passband by \citet{ESA1997} is shown as the dashed line for comparison. A significant difference with respect to the \citet{ESA1997} passband has been found, as reported before \citep{BessellMurphy}. The lower two panels in Fig.~\ref{Fig:3} show the relative flux residuals (defined as observed minus calculated over observed flux) as a function of the HIPPARCOS magnitude and Johnson $B-V$ colour for the 206 calibration sources, respectively. No systematic trends with magnitude or colour can be observed. The standard deviation, indicated by the highlighted regions in Fig.~\ref{Fig:3}, is $\rm 1.85\%$.\par
\begin{figure}
   \centering
   \includegraphics[width=0.5\textwidth]{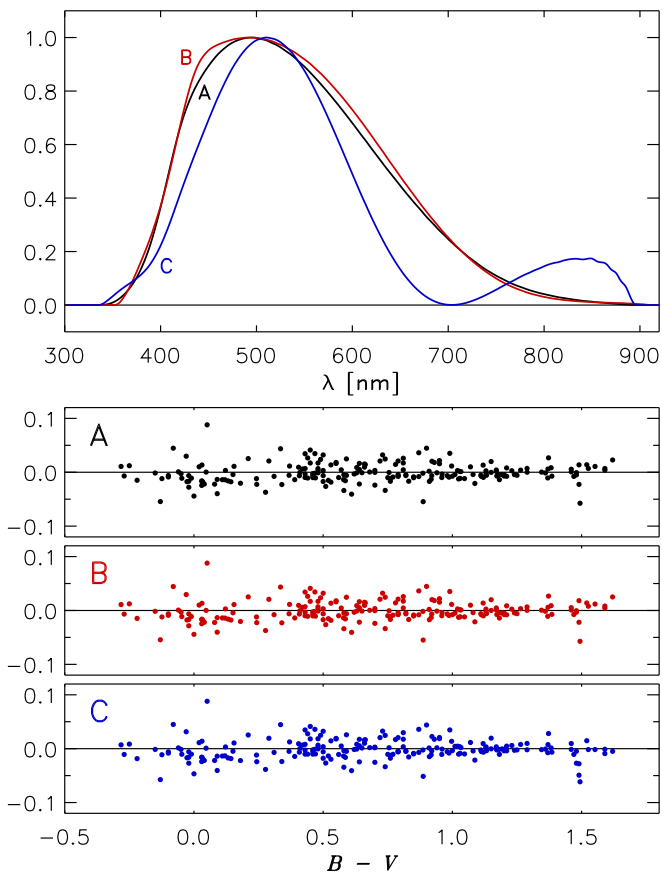}
   \caption{Comparison of three (of the infinitely many) possible passband solutions for HIPPARCOS. The black curve (A) in the upper panel is the same as in Fig.~\ref{Fig:3}. The three lower panels show the relative flux residuals for the three curves (A, B, C) versus the Johnson $B-V$ colour.
              \label{Fig:4}}
    \end{figure}

To illustrate the uncertainty introduced by the unconstrained orthogonal component, Fig.~\ref{Fig:4} shows three solutions for the HIPPARCOS passband in comparison. All three components have the same parallel component, and only differ in $p_\perp$. Thus, they result in essentially the same relative residuals for the 206 calibration sources, as shown in the lower three panels of Fig.~\ref{Fig:4}. Solution A in this figure is the passband selected as the ''preferred'' solution and shown in Fig.~\ref{Fig:3}. Passband B is rather similar to solution A, derived from slightly shifting $p_{ini}$ in the wavelength axis. Solution C has been chosen strongly different from solution A by making use of the null space approach, with one degree of freedom. Although all three solutions are indistinguishable from the set of calibration spectra used, a solution so grotesque as solution C may safely be rejected based on the a-priori knowledge on the passband. The solutions A and B however are already similar enough to make a decision about which one is closer to the true passband difficult. The implications of the fundamental ignorance about the correct orthogonal component is discussed in more detail in Sec. \ref{sec:orteffect}.

\subsection{Tycho $B_T$ passband \label{sec:BT}}

\begin{figure}
   \centering
   \includegraphics[width=0.5\textwidth]{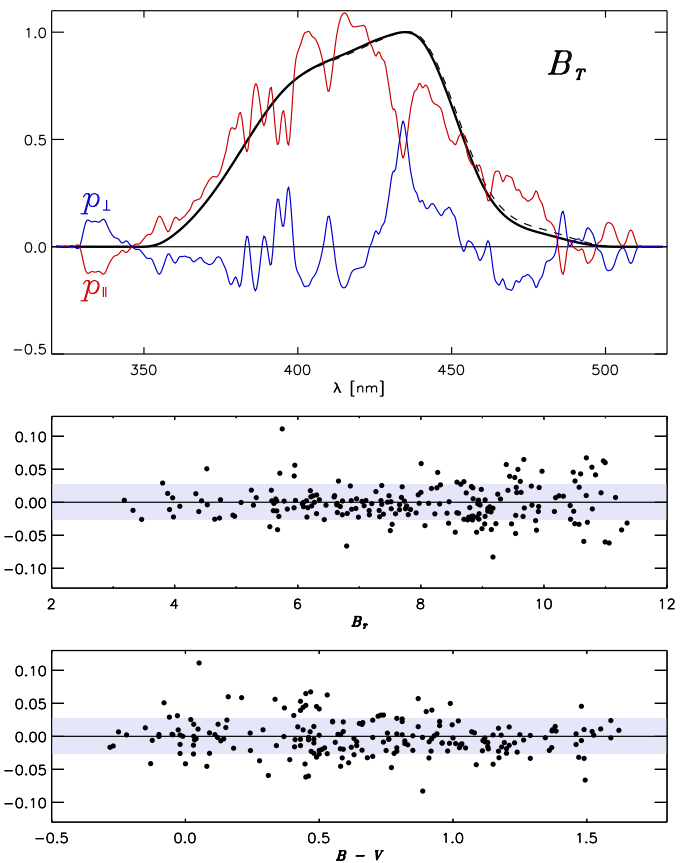}
   \caption{As Fig.~\ref{Fig:3}, but for $B_T$.
              \label{Fig:5}}
    \end{figure}

The $B_T$ passband has been determined from 210 NGSL spectra. As for HIPPARCOS, the solution of Eq. (\ref{eq:7}) for the parallel component was modified, using the random sampling approach and a polynomial of degree 3. As the target passband $p_{ini}$, the passband by \citet{ESA1997} for $B_T$ was used. In this case, a solution very close to the one by \citet{ESA1997} was found. The passband with its parallel and orthogonal components and the relative residuals versus $B_T$ magnitude and $B-V$ colour are shown in Fig.~\ref{Fig:5}. Again, no trend in residuals with magnitude and colour is visible. As compared to the case of the $H_p$ passband, an increase in the residuals with increasing magnitude can be observed. This effect can be understood by the larger random error on the observed fluxes of the calibration stars, as compared to the HIPPARCOS case. For faint stars, this random error becomes dominant over the uncertainty in the NGSL spectra, causing the increase in residuals. The standard deviation of the relative flux residuals is $\rm 2.70\%$.

\subsection{Tycho $V_T$ passband}

\begin{figure}
   \centering
   \includegraphics[width=0.5\textwidth]{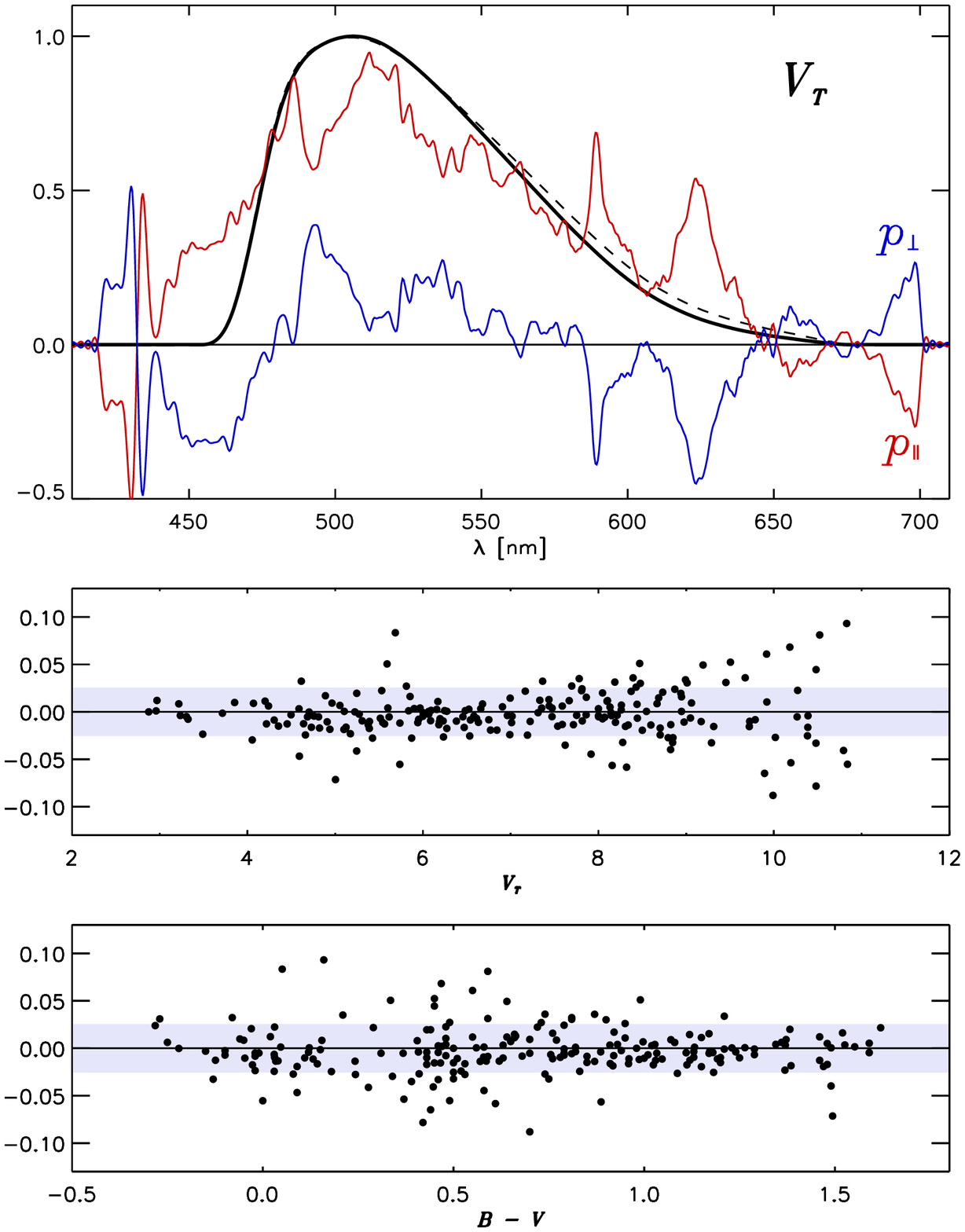}
   \caption{As Fig.~\ref{Fig:3}, but for $V_T$.
              \label{Fig:6}}
    \end{figure}

The $V_T$ passband has been derived analogously to $B_T$ from 210 NGSL spectra, with the only difference of using a polynomial of degree 2 for the modification of the passband by \citet{ESA1997}. Again, a solution very close to the passband by \citet{ESA1997} was found, only being slightly lower at longer wavelengths. The passband solution with the parallel and orthogonal component and the relative residuals are presented in Fig.~\ref{Fig:6}. No trends in residuals with magnitude and colour are visible. The increase in residuals with increasing magnitude, like in the case of $B_T$, is present for $V_T$ as well. The standard deviation of the relative flux residuals is 2.56\%.

\subsection{Gaia DR1 $G$ passband \label{sec:Gaia}}

\begin{figure}
   \centering
   \includegraphics[width=0.5\textwidth]{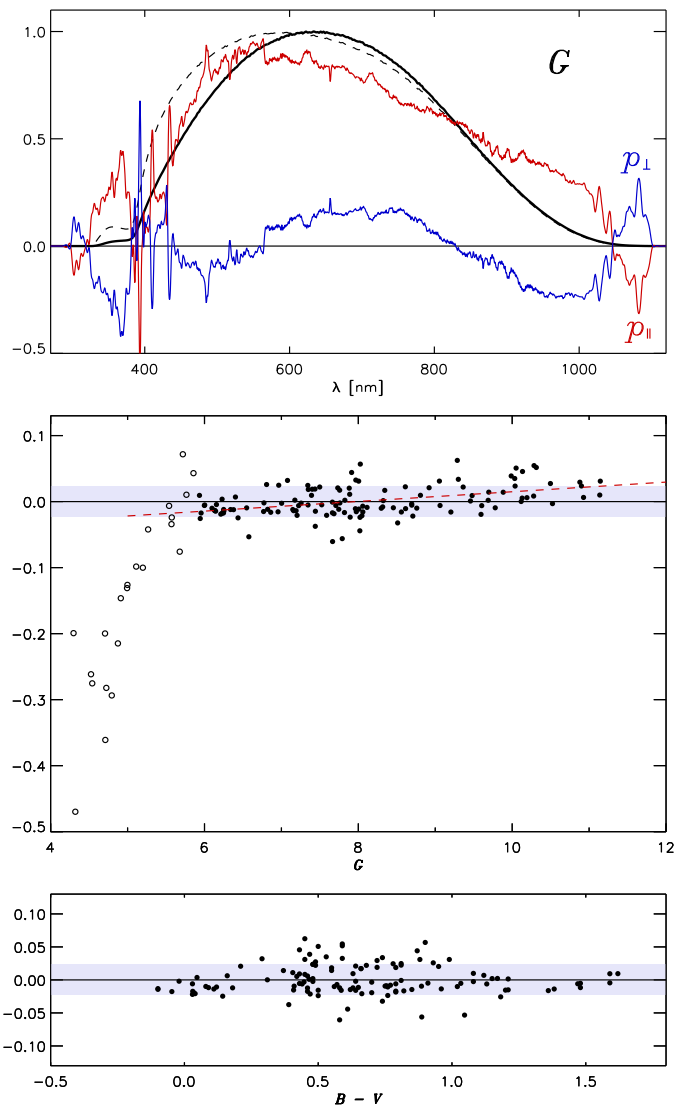}
   \caption{As Fig.~\ref{Fig:3}, but for $G$. The open symbols in the central panel are for the calibration stars brighter than $G = {\rm 5.9}$, which have been excluded from the passband determination. The dashed red line shows a possible linear trend in residuals with $G$ magnitude.
              \label{Fig:7}}
    \end{figure}

A set of 124 NGSL spectra was used here for deriving the passband of {\it Gaia} DR1 $G$. The procedure applied for $G$ differs somewhat from the cases of the other passbands. As the passband to be approximated, $p_{ini}$, the nominal $G$ passband \citep{Jordi2010} was used. No solution close to the nominal passband could be found in the random sampling approach, indicating a significant difference between the nominal and the actual $G$ passband. Such a difference is consistent with a wavelength dependency of contamination effects reported to affect the {\it Gaia} photometry \citep{Gaia2016a}. The transmission loss due to contamination is colour dependent, indicating a larger loss at shorter wavelengths (see Gaia DR1 online documentation\footnote{\url{https://gaia.esac.esa.int/documentation/GDR1/index.html}}). We therefore decided to use as the preferred solution for the $G$ passband a solution that differs from the nominal passband mainly at short wavelengths. Thus, we were using the solution for the parallel component of the passband obtained from Eq. (\ref{eq:7}), and used the null space approach to find a solution for the passband with most deviations from the nominal passband at short wavelengths. A polynomial of degree 3, with one degree of freedom, was used in this process. The passband solution, with the parallel and orthogonal components, and the relative residuals, are shown in Fig.~\ref{Fig:7}.\par
Regarding the magnitude dependency of the residuals, we observed a very strong systematic deviation in the residuals, for sources brighter than about 5.9 magnitudes in $G$, as previously reported by \citet{Maiz2017}. The residuals for these sources are shown as open symbols in the central panel of Fig.~\ref{Fig:7}, and these sources have been excluded from the passband determination. The onset of the strong trend at about 5.9 mag is probably caused by saturation effects in the {\it Gaia} photometry.\par
Even after excluding the flawed bright sources from the passband determination, a tentative slight magnitude dependency of the residuals may be present. The trend is indicated by the dashed line in the central panel of Fig.~\ref{Fig:7}, representing a linear fit to the relative residuals. This trend, amounting to roughly 0.7\% per magnitude drift, however remains close to the limit of accuracy reached within this work. An increase in the scatter of the residuals with increasing magnitude is absent for $G$, indicating a lower random error on the observed fluxes than for $B_T$ and $V_T$. The standard deviation of the relative residuals is 2.29\%.

\subsection{Relations between passbands}

\begin{figure}
   \centering
   \includegraphics[width=0.5\textwidth]{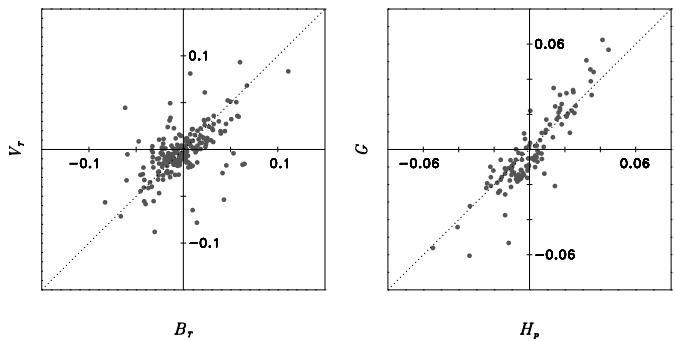}
   \caption{Relative flux residuals in the calibration of the $V_T$ passbands versus the corresponding residuals for $B_T$ (left panel), and the residuals for $G$ versus the residuals for $H_p$ (right panel), for the calibration SPDs common in the two cases.
              \label{Fig:8}}
    \end{figure}

As discussed in Sec. \ref{sec:uncertainties}, we expect the residuals in synthetic photometry being dominated by the uncertainty in the SPDs used for the calibration of the passbands. As there is a large overlap between the sets of calibration sources for the four different passbands, we see a strong correlation between their residuals. This is illustrated for two examples in Fig.~\ref{Fig:8}, showing the residuals for $V_T$ versus the residuals for $B_T$, and the residuals for $G$ versus $H_p$, for the sources common in the calibration of the passbands. For the case of $V_T$ versus $B_T$, we see a larger scatter and a lower correlation as compared to $G$ versus $H_p$, as in the first case, the random error on the observed photometry still is significant compared to the error on the calibration SPDs, in particular for the faint sources. For the $G$ versus $H_p$ case, the scatter is smaller and the correlation is larger. For the photometry of HIPPARCOS and {\it Gaia}, the errors in the calibration SPDs dominate the residuals.\par
\begin{figure}
   \centering
   \includegraphics[width=0.5\textwidth]{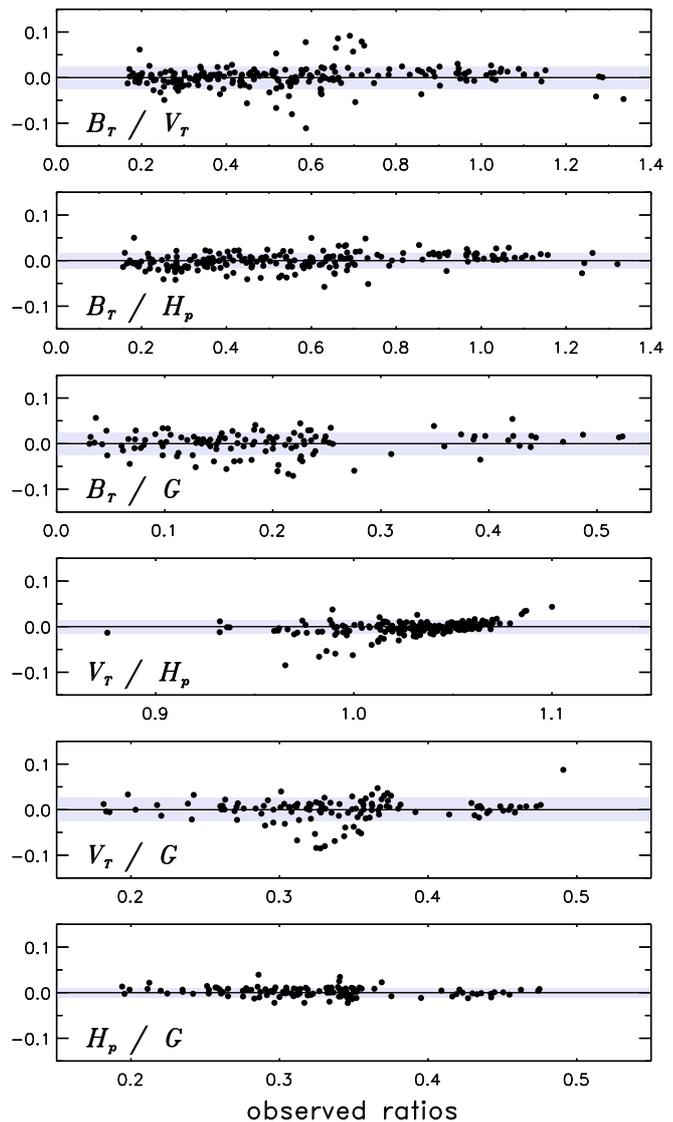}
   \caption{Relative residuals in the ratios of fluxes for the 6 combinations of passbands, versus the observed flux ratios.
              \label{Fig:9}}
    \end{figure}

The domination of the error on the calibration SPDs makes it, strictly speaking, necessary not only to minimise the norm of the residual vector on the photometry, i.e. $\bf c$ in Eq. (\ref{eq:7}), but also the norm of the residuals of the matrix $\bf A$. As a stringent treatment of this problem requires a reliable error model on $\bf A$ which is, as discussed in Sec. \ref{sec:uncertainties}, not available, we nevertheless determine the passband by minimising the norm of the residuals on $\bf c$. We may confirm that the solution found this way is meaningful by not only considering the residuals for each passband separately, but also ensuring that the residuals of the ratios of the fluxes between pairs of passbands are free of trend with colour. These residuals are shown in Fig.~\ref{Fig:9} for the passbands derived in this work. In general, the residuals in the ratios show no dependency on the ratio of fluxes, indicating that the passbands found do not only provide a good reproduction of the observed fluxes, but also of the observed colours. Only in the case of $B_T / H_p$, a slight systematic increase of the residuals in the range of ratios between 0.7 and 1.2 might be detected. In particular for the case of of $H_p / G$, where the errors on the observed photometry play the least role, the influence of the correlation of the errors can be seen by the reduction of the relative error on the ratio of fluxes, as compared to the relative errors on the fluxes individually (cf. Figs. \ref{Fig:3} and \ref{Fig:7}).

\section{Comparison with other results \label{sec:compWithOthers}}
In this section, we compare the passband solutions obtained in this work with previously published passbands for $H_p$, $B_T$, $V_T$, and $G$. For the HIPPARCOS and Tycho passbands, we compare our solutions with the solutions provided by \citet{ESA1997}, \citet{Bessell2000}, and \citet{BessellMurphy}. For the {\it Gaia} $G$ passband, we compare with the nominal pre-launch {\it Gaia} passband \citep{Jordi2010} and the passband by \citet{Maiz2017}.\par
Different authors have used different sets of calibration SPDs. \citet{Bessell2000} has used Vilnius spectra \citep{Vilnius} for the passband determination, while \citet{BessellMurphy} have used NGSL and MILES spectra. \citet{Maiz2017} has used a compilation of NGSL and CALSPEC \citep{Bohlin2017} spectra. Also, different selection criteria were applied when choosing spectra for calibration, as far as magnitudes and variability are concerned. As the shape of the passband $p$ is, as previously shown, subject to considerable uncertainties due to the unconstrained orthogonal component, we are not only comparing the passbands themselves, but also the projection of the passbands onto the subspace spanned by the calibration sources used in this work, i.e. we are comparing the parallel components $p_\parallel$ with respect to our calibration set. In the later case, we are able to identify ''substantial'' differences in the passband, instead of differences that only affect the unconstrained component. While the differences in the passband $p$ could be caused by simply obtaining different estimates for the orthogonal component $p_\perp$ in the process of determining the passbands, there are two effects that could cause differences in $p_\parallel$. First, the set of calibration sources used by other authors may not span the same subspace as the calibration sources used in this work. The other passbands are then optimised for a different subspace, and, if this subspace does not fully contain the subspace used in this work, one may find a suboptimal solution compared to the solution of this work, constrained to the subspace used in this work. Second, the method of modifying the initial passband used in previous publications may not allow to find the correct parallel component of the passband. In this case, one may obtain an improved, but not optimal solution for the passband. Both effects may play a role in the comparison of the results of this work with other passband solutions.\par
The parallel passband components compared in this section always refer to the normalised passbands, and the normalisation factor of a passband depends on the sum of the parallel and the orthogonal component. It therefore has to be kept in mind that different scaling factors may apply to the parallel components from different solutions for the same passband.

\subsection{HIPPARCOS passband}

\begin{figure}
   \centering
   \includegraphics[width=0.5\textwidth]{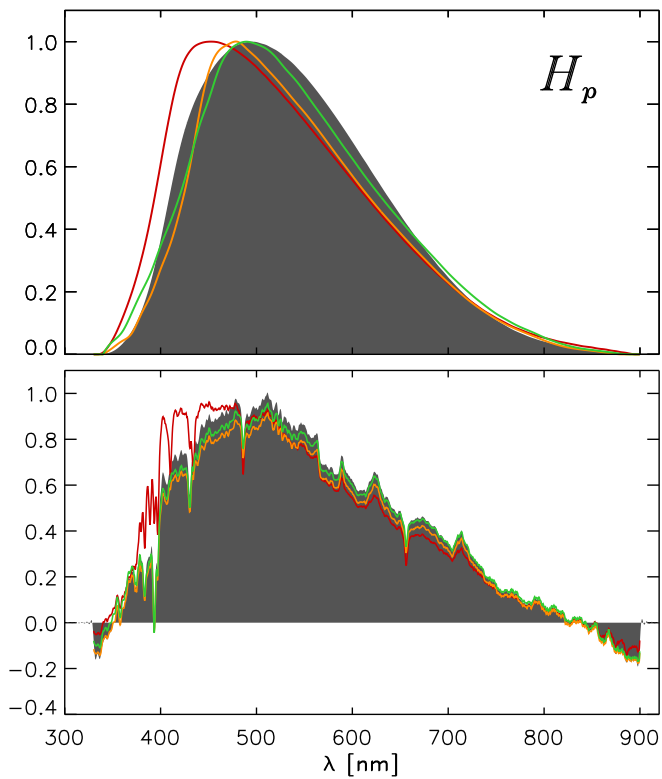}
   \caption{Upper panel: HIPPARCOS passbands from different publications. Lower panel: projection of the passbands to the subspace spanned by the calibration sources of this work. Grey shaded regions: Solution from this work. Red curves: \citet{ESA1997}. Orange curves: \citet{Bessell2000}. Green curves: \citet{BessellMurphy}.
              \label{Fig:11}}
    \end{figure}

The $H_p$ passband of this work is compared with other passbands in Fig.~\ref{Fig:11}, upper panel, while the corresponding parallel components are compared in the lower panel. Clear differences between the four different passbands can be seen. The passband by \citet{ESA1997} differs most, having a much higher response at short wavelengths. The other three passbands are more similar, the solution of this work being somewhat broader than the solutions by \citet{Bessell2000} and \citet{BessellMurphy}. If the passband is restricted to the parallel component, the differences become smaller. For \citet{ESA1997}, there still is a substantial difference with respect to the other parallel components, again showing a higher response at short wavelengths. The parallel components of the passbands by \citet{Bessell2000} and \citet{BessellMurphy} differ only little, although the passbands are clearly different. This indicates that the change between \citet{Bessell2000} and \citet{BessellMurphy} mainly affect the unconstrained orthogonal component, at least as the calibration sources used in this work are concerned. Both parallel components are also similar to the parallel component derived in this work. For wavelengths larger than about 600 nm, the agreement with the result of this work is very good. For wavelengths between about 400 and 600 nm, however, we derive a slightly higher response in this work.\par
The standard deviations of the relative flux residuals is 4.69\% for the \citet{ESA1997} passband, 1.85\% for the \citet{Bessell2000} passband, and 1.91\% for the \citet{BessellMurphy} passband, as compared to 1.85\% for the passband derived in this work. Thus, the \citet{ESA1997} passband is clearly not optimal, while the differences among the other three passbands are very small. Also the dependency of the residuals from colour is rather similar for the last three passbands. They thus differ strongly in their orthogonal components, resulting in very different shapes, but their differences in the parallel component are sufficiently small to be within the uncertainties introduced by the calibration data of this work.\par
With respect to the set of calibration sources used in this work, the passbands by \citet{Bessell2000}, \citet{BessellMurphy}, and this work, result in similar goodnesses-of-fit, and are thus essentially equivalent.

\subsection{Tycho passbands}

\begin{figure}
   \centering
   \includegraphics[width=0.5\textwidth]{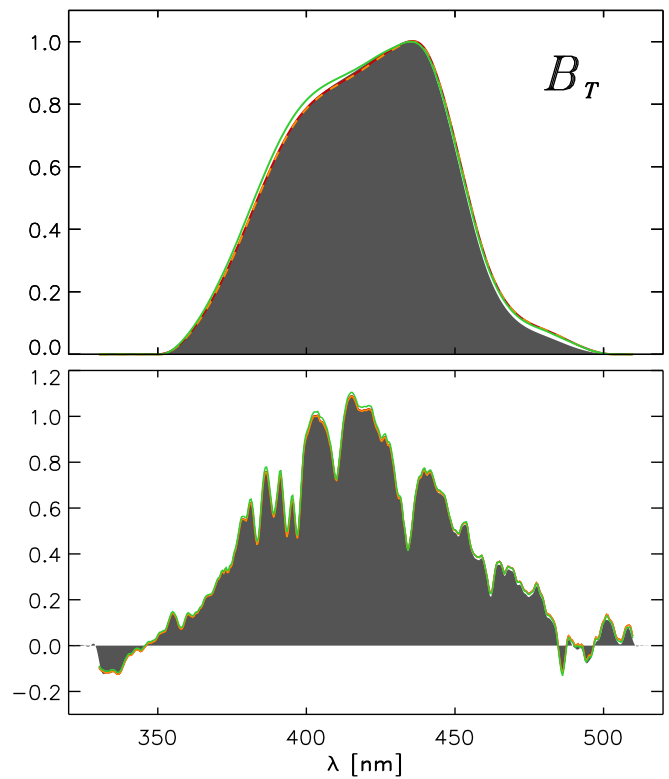}
   \caption{As Fig.~\ref{Fig:11}, but for the $B_T$ passband.
              \label{Fig:12}}
    \end{figure}

\begin{figure}
   \centering
   \includegraphics[width=0.5\textwidth]{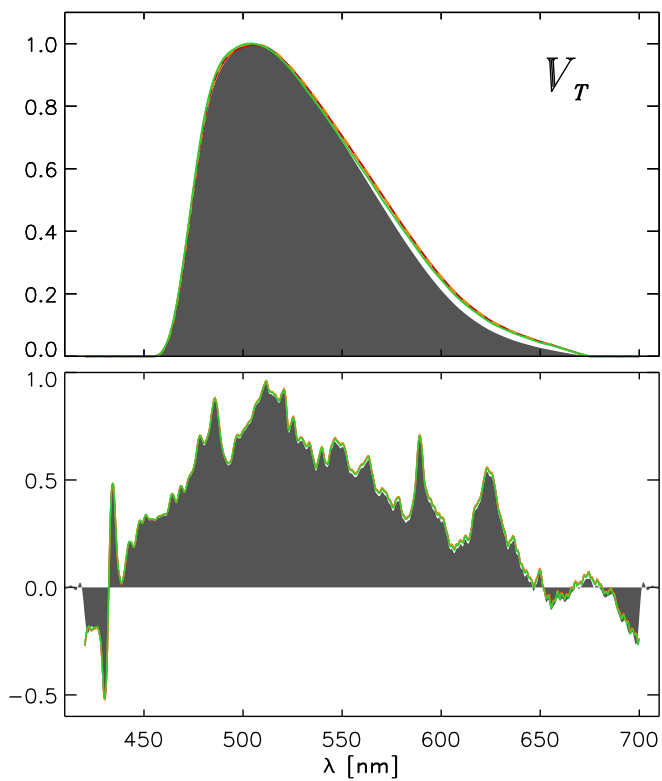}
   \caption{As Fig.~\ref{Fig:11}, but for the $V_T$ passband.
              \label{Fig:13}}
    \end{figure}

Comparisons of  different solutions for the $B_T$ and $V_T$ passbands are shown in Fig.~\ref{Fig:12} and \ref{Fig:13}, respectively. For $B_T$, all the passbands as well as their parallel components, are very similar. At longer wavelengths, between about 450 and 500 nm, the passband of this work has slightly lower responses than the others, while at wavelengths around 400 nm, the solution by \citet{BessellMurphy} has slightly higher response than the others. When considering the parallel component, these differences become even less prominent.\par
For the $V_T$ passband, we also find a lower response at long wavelengths, while the passbands from other publications agree very well among each other. In the parallel component, the differences are rather small, however. \par
The standard deviation of the relative flux residuals for the four Tycho $B_T$ passbands compared are 2.88\% for \citet{ESA1997} and \citet{Bessell2000}, 2.74\% for \citet{BessellMurphy}, and 2.70\% for this work. For the Tycho $V_T$ passband, the values are 2.65\% for \citet{ESA1997} and \citet{Bessell2000}, 2.60\% for \citet{BessellMurphy}, as compared to 2.56\% for the passband of this work. Thus, we may therefore take all Tycho passbands as essentially equivalent, as far as the subspace of ${\mathcal L}^2$ spanned by the calibration sources of this work is concerned.

\subsection{Gaia DR1 passband}

\begin{figure}
   \centering
   \includegraphics[width=0.5\textwidth]{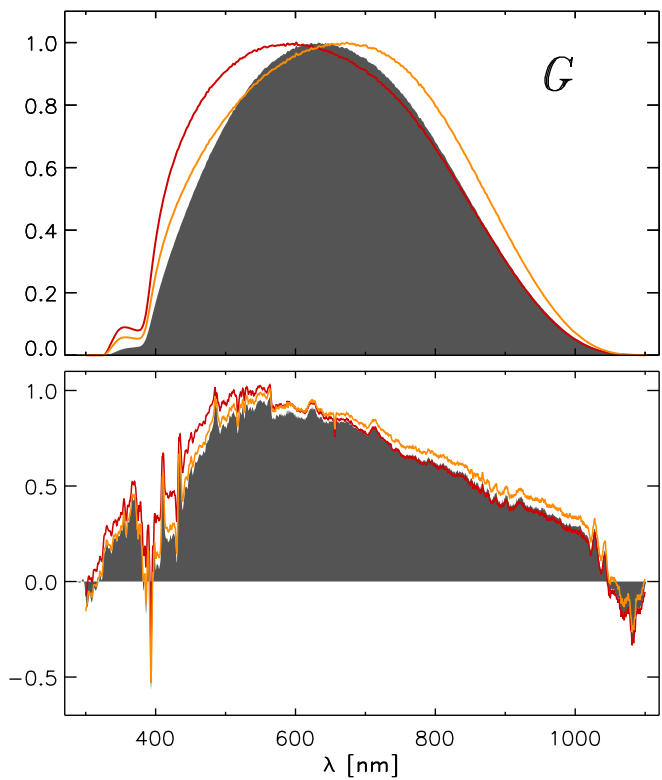}
   \caption{As Fig.~\ref{Fig:11}, but for {\it Gaia} DR1 passband. Grey shaded regions: Solution from this work. Red curves: \citet{Jordi2010}. Orange curves: \citet{Maiz2017}.
              \label{Fig:14}}
    \end{figure}

For the {\it Gaia} DR1 passband, shown in Fig.~\ref{Fig:14}, we see rather strong differences in shape between the nominal passband, the solution by \citet{Maiz2017}, and this work. Comparing the result of this work with the pre-launch passband by \citet{Jordi2010}, we find good agreement at wavelengths larger than about 700 nm, while at shorter wavelengths we find a lower response. This may indicate that the response loss observed in the {\it Gaia} observations due to contamination is indeed stronger at short wavelengths, an assumption that has already been used when estimating the orthogonal component of the $G$ passband, as mentioned in Sec. \ref{sec:Gaia}.\par
The parallel component of the passband by \citet{Maiz2017} is very close to the parallel component of this work at short wavelengths, but having a higher transmissivity than the result of this work and the nominal passband at longer wavelengths. As \citet{Maiz2017} uses NGSL spectra for the passband calibration, as done in this work, one may expect both solutions for the $G$ passband being determined on a similar subspace of ${\mathcal L}^2$. However, \citet{Maiz2017} modifies the nominal passband by multiplying the nominal passband with a single parameter power law. It may be possible that this modification approach is not able to produce the optimal parallel component. The obtained solution of the passband then may provide an improvement to the nominal passband, but the correction of the nominal passband is overdone at large wavelengths, resulting in a too high response.\par
The standard deviations of the relative flux residuals for the pre-launch $G$ passband by \citet{Jordi2010} is 4.22\%, as compared to 2.32\% for the passband by \citet{Maiz2017} and 2.29\% for this work. The pre-launch passband estimate is thus poor for {\it Gaia} DR1. The solution by \citet{Maiz2017} and this work are again very similar in the level of accuracy with which the photometry of the calibration sources can be reproduced. Also the dependency of the relative flux residuals on colour is similar in both cases. As far as the calibration sources used in this work are concerned, the passbands by \citet{Maiz2017} and this work are thus essentially equivalent; they differ strongly in $p_\perp(\lambda)$, but the differences in $p_\parallel(\lambda)$ are within the uncertainties.

\section{Effects of guessing $p_\perp(\lambda)$ \label{sec:orteffect}}

As worked out in Sec. \ref{sec:formulation} and illustrated for the HIPPARCOS passband in Sec. \ref{sec:hipparcos}, a set of calibration spectra together with corresponding photometric observations does not fully constrain a passband. Even taking physical constraints to the passband, that is, smoothness and non-negativity, into account, a considerable uncertainty in the shape of the passband remains. In practice, the uncertainty in shape has relatively little impact on the synthetic photometry. Stellar SPDs vary in rather tight bounds, and with a few basis functions, a large fraction of spectra can be well represented. This may be compared to the fact that photometric measurements in a few passbands already allows for a meaningful stellar classification. Thus, if the set of calibration spectra covers several spectral classes, the resulting parallel component of the passband is sufficient for a reliable synthetic photometry for a wide range of astronomical objects. The uncertainty in the orthogonal component, although allowing for visually impressive variations in the full passband, as illustrated in Fig.~\ref{Fig:4}, is of little to no relevance for the synthetic photometry of these SPDs. The choice of the orthogonal component has significant impact only for SPDs which strongly differ from the SPDs of the set of calibration sources. Strongly different means here, that a SPD cannot be well represented by a linear combination of the calibration SPDs, or, stated in the functional analytic formalism of this work, that the SPD has a significant component outside the subspace spanned by the calibration SPDs. An example for such SPDs may be cool M-type stars, as their SPDs differ by their Planck function and strong absorption features very much from the SPDs of hotter stars. At the same time, the frequent variability of M-type stars makes it difficult to include their SPDs in a calibration set for passband determination. Also strongly non-stellar SPDs with large variability in shape, as is e.g. the case for quasi-stellar objects, can be expected to be among the SPDs for which the choice of the orthogonal component of the passband is relevant.\par
As the only way to remove the uncertainty in the orthogonal component is the (in practice unfeasible) extension of the set of calibration SPDs such that all astronomical SPDs can be well approximated by a linear combination of the calibration SPDs, it is desirable to have a measure at hand which allows to quantify the degree to which any given SPDs is sensitive to the uncertainty. For a stringent determination of the component of a SPDs orthogonal to the set of calibration sources, the basis functions used in the passband determination are required. For practical purposes, a simpler and more convenient measure for the sensitivity of a SPD for the choice of $p_\perp$, may be sufficient.\par
As a simplified measure, we may chose the ratio between the flux of some SPD $s(\lambda)$ in the orthogonal passband component and in the parallel passband component, i.e. $\langle\, p_\perp\, | \, s \, \rangle / \langle\, p_\parallel\, | \, s \, \rangle$. This ratio is in the range $[0,\infty)$, with zero in the case that the flux in the orthogonal component is zero, and all synthetic photometry arises from the parallel component, and infinity in the reverse case. We may obtain a somewhat more convenient range and a more intuitive interpretation if we slightly modify the measure by dividing the fluxes by the $l_2$-norm of $p_\perp$ and $p_\parallel$, respectively, and taking the inverse tangents of the resulting quantity. If we name the result $\gamma$, we have
\begin{equation}
\gamma = {\rm atan}\left( \,\left| \,  \frac{\langle\, p_\perp\,|\, s\, \rangle}{||p_\perp ||_2} \frac{||\,p_\parallel\,||_2}{\langle\, p_\parallel\,|\,s\, \rangle} \, \right| \, \right) \quad . \label{eq:gamma}
\end{equation}
The $l_2$-norm of some function $p$ is simply $|| \, p \, ||_2 = \sqrt{\langle\, p \, | \, p \, \rangle}$. As the terms $\langle\, p_\perp\,|\, s\, \rangle\,  ||p_\perp||_2^{-1}$ and $\langle\, p_\parallel\,|\, s\, \rangle\,  ||p_\parallel||_2^{-1}$ are the projections of the SPD $s(\lambda)$ onto unit vectors in the directions of the orthogonal and parallel passband components, respectively, the inverse tangent of the ratio of these two quantities is the angle between the SPD $s(\lambda)$ and the parallel component of the passband, measured in the plane spanned by $p_\perp$ and $p_\parallel$, and ranging between $[0^{\rm o},90^{\rm o}]$. Taking the absolute value in the argument of the inverse tangent in Eq. (\ref{eq:gamma}) is introduced as we are not interested in the orientation of the projection. Thus, we obtain a simple graphical interpretation of the quantity $\gamma$. If, for some SPD, $\gamma=0$, it is aligned with the parallel component of the passband, and the guess for the orthogonal component is irrelevant for that SPD. If the angle $\gamma$ increases, the SPD turns away from the parallel component, and the orthogonal component is contributing more and more to the synthetic photometry. For $\gamma=90^{\rm o}$, the SPD is aligned with the orthogonal component, and the synthetic photometry depends entirely on the guessed function $p_\perp$, and not at all on the constrained function $p_\parallel$. It has to be kept in mind that the angle $\gamma$ is measured inside the plane spanned by $p_\perp$ and $p_\parallel$, and thus depend on the estimate of $p_\perp$ itself. If however the orthogonal component is not strongly wrong, a case which the a-priori knowledge and the physical constraints on the passband shall prevent, in practice the quantity $\gamma$ may be sufficient to estimate the degree to which a given SPD is sensitive to the unconstrained part of a passband. What is required for the computation of $\gamma$ is the parallel and orthogonal components of the passband, and their $l_2$ norms, which are listed in Table~\ref{tab:2} for the four passbands provided in this work. With these two components at hand, one can compute the synthetic photometry for a SPD by solving two integrals instead of one (i.e., using Eq. (\ref{eq:passbandDecomposition})), and by doing so getting the measure $\gamma$.\par
\begin{figure}
   \centering
   \includegraphics[width=0.5\textwidth]{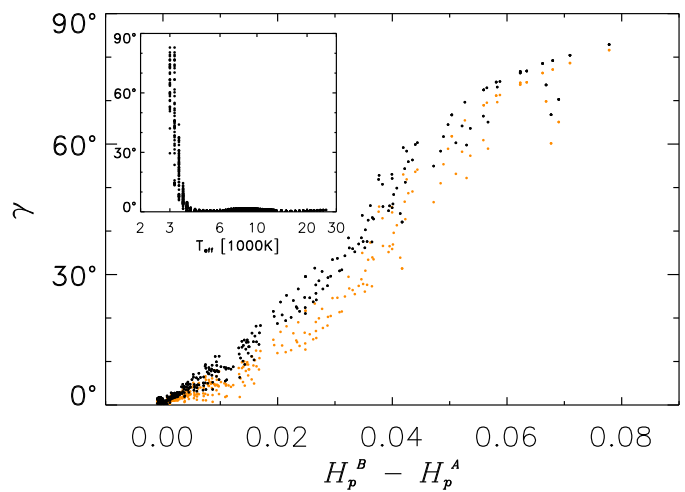}
   \caption{The angle $\gamma$ versus the difference in magnitude between the HIPPARCOS passbands A and B for the stellar spectra by \citet{Coelho2014}. Black dots are $\gamma$ values computed with passband A, orange dots computed with passband B. The inset shows $\gamma$ versus the effective temperature of the star.
              \label{Fig:10}}
    \end{figure}

To illustrate the effects of the choice of $p_\perp$ and the usage of $\gamma$, we produce an example based on a set of synthetic stellar spectra, namely the set of 3727 spectra by \citet{Coelho2014}. These synthetic stellar spectra with a high spectral resolution cover a range of effective temperatures from 3000~K to 26000~K, with log($g$) in the range of $\rm -0.5$ to 5.5, and with different chemical mixtures. We compute the $H_p$ magnitudes from these spectra, using two solutions for the HIPPARCOS passband which differ only in the choice of $p_\perp$. The two $H_p$ passbands are our preferred passband solution shown in Fig.~\ref{Fig:3} and as passband A in Fig.~\ref{Fig:4}, and the similar passband B as shown in Fig.~\ref{Fig:4}. The difference between the magnitudes resulting from both HIPPARCOS passbands for each spectrum, $H_p^B - H_p^A$, illustrates the influence of the choice of the passband component orthogonal to the set of calibration spectra. In Fig.~\ref{Fig:10}, the angle $\gamma$  is plotted against the difference $H_p^B - H_p^A$ for all synthetic spectra in the set by \citet{Coelho2014}. $\gamma$ is computed for both the $H_p$ solutions A and B, to illustrate the dependency on the choice of $p_\perp$. The difference in calculated magnitudes correlates strongly with the value for $\gamma$. For the majority of the synthetic spectra, the difference $H_p^B - H_p^A$ is essentially negligible. The values of the angle $\gamma$ are close to zero for these sources. For some synthetic spectra, there is however a clear and systematic increase in $H_p^B - H_p^A$, which is associated with an increase of $\gamma$. These sources are the spectra with effective temperatures at the lower end of the range covered, as can be seen from the inset in Fig.~\ref{Fig:10}, displaying the relationship between $\gamma$ and the effective temperature. Thus, SPDs from \citet{Coelho2014} corresponding to effective temperatures larger than about 4000K up to the limit of 26000K are well represented by the set of NGSL spectra used to calibrate the HIPPARCOS passband. Their computed magnitudes depend essentially on the parallel component of the $H_p$ passband, and are therefore insensitive to the estimate used for the orthogonal component. Their corresponding $\gamma$ angles are consequently rather small. For SPDs corresponding to effective temperatures less than about 4000K, no good approximation by the set of NGSL spectra is possible. These SPDs have a significant component orthogonal to all calibration SPDs, and their calculated magnitudes depend on the choice of $p_\perp$. The angle $\gamma$ increases for these SPDs, reaching values over $\rm 80^o$ in the extreme cases. For these SPDs, the calculated magnitudes are dominated by the estimated component $p_\perp$, and not by the constrained parallel component $p_\parallel$. Changing $p_\perp$ therefore results in a systematic shift in the magnitude predicted from the passband. In the example shown in Fig.~\ref{Fig:10}, the difference in the two $H_p$ passbands amounts to a shift up to about 80 mmag for the coolest sources. It should noted that the value of this difference depends on the choices of the passband solutions compared, and the SPDs considered, and therefore does not allow for any generalisations.\par
The general behaviour of $\gamma$ with the shift in magnitude is the same if $\gamma$ is computed from solution A or solution B. The angle $\gamma$ may thus be used to quantify the degree to which a given SPD is sensitive to systematic errors introduced by the uncertainty related to $p_\perp$. It can however not make any statement about if or how much the calculated magnitude for the SPD is differing from the true value. The actual systematic error depends on the difference between the estimated and the true $p_\perp$, while the later is in principle unknown in a situation that requires a passband calibration. Nevertheless, $\gamma$ may be a useful indicator for the reliability of synthetic photometry. A small value of a few degrees or less indicates that the calculated magnitude relies on the well-known component of the passband, while a large value of $\gamma$ indicates a strong dependency on the guess for $p_\perp$. Results based on synthetic photometry associated with large values of $\gamma$ may therefore include the possibility of systematic effects due to passband uncertainties in their discussion.

\section{Zero points \label{sec:zeropoints}}

We compute the zero points ($zp$) for the passbands derived in this work in the VEGAMAG system, i.e. we apply Eq. (\ref{eq:magnitude}), with
\begin{equation}
zp = 0.03 + 2.5\cdot {\rm log_{10}}\left( \bar{c}_{\alpha {\rm Lyr}} \right) \quad .
\end{equation}
Here, $\bar{c}_{\alpha {\rm Lyr}}$ denotes the photometry of Vega in the passband considered, and the value of 0.03 corresponds to the assumption of $V {\rm = 0.03}$ for Vega \citep{Jordi2010}. We computed $\bar{c}_{\alpha {\rm Lyr}}$ using the CALSPEC spectrum {\tt alpha\_lyr\_stis\_008} for the SED of Vega.\par
Note that the zero point for the {\it Gaia} $G$ passband applies to the published DR1 electron counts in electrons per second \citep{Carrasco2016}, recorded within the {\it Gaia} telescope aperture of 0.7278~m$^2$ \citep{Gaia2016a}, and not within the normal area of $\rm 1~m^2$.\par
We also compute the zero points in the AB magnitude system, where the zero point is given by
\begin{equation}
zp = 2.5\cdot {\rm log_{10}}\left(\bar{c}_J\right) \quad,
\end{equation}
with $\bar{c}_J$ the number of photons detected in the passband for a SED with 3631~Jy at all frequencies \citep{OkeGunn}.\par
The computed zero points for the four passbands considered in this work are listed in Table~\ref{tab:2}, together with the mean wavelengths $\lambda_m$ of the passbands,
\begin{equation}
\lambda_m \equiv \frac{\int_{\lambda_0}^{\lambda_1}\, \lambda \cdot p(\lambda)\, {\rm d}\lambda} {\int_{\lambda_0}^{\lambda_1}\, p(\lambda)\,{\rm d}\lambda} \quad ,
\end{equation}
and the pivot wavelengths $\lambda_p$ \citep{Koornneef},
\begin{equation}
\lambda_p \equiv \sqrt{\frac{\int_{\lambda_0}^{\lambda_1}\, \lambda \cdot p(\lambda)\,{\rm d}\lambda}{\int_{\lambda_0}^{\lambda_1}\, \lambda^{-1}\cdot p(\lambda)\,{\rm d}\lambda}} \quad .
\end{equation}
The zero points the Table~\ref{tab:2} are valid for passbands normalised to a maximum value of one. For $G$, this zero point is 0.351 magnitudes larger than the one for the unnormalised passband. As for the $G$ passband, a tentative dependency of the residuals on $G$ magnitude has been observed in this work, we shall note that the zero point for $G$ is valid for a magnitude of $G$~=~8.\par
For the reference SPDs in both photometric systems, the VEGAMAG and the AB, we can compute the angle $\gamma$ to estimate the influence of the choice of the orthogonal component of the passband onto the zero points. For the case of Vega, we obtain values of 1 degree or less for all four passbands, indicating that the SPD of Vega is well represented by the calibration sources used in this work, and consequently that the zero points are hardly affected by the choice of $p_\perp(\lambda)$. For the AB system, $\gamma$ is below 1.1 degrees for $H_p$, $B_T$, and $V_T$, indicating little influence of the choice of $p_\perp(\lambda)$. Only for the $G$ passband, $\gamma$ reaches $\rm 7^o$, thus indicating a sensitivity of the zero point on the choice of $p_\perp(\lambda)$.\par
Although a reliable computation of the zero points is possible as far as the choice of the calibration sources is concerned, we nevertheless introduce a dependency on the orthogonal component by the normalisation of the passbands. We follow the convention of presenting passbands normalised to the maximum value in this work, and the normalisation factor depends on the full passband $p$, not on the parallel component $p_\parallel$ only. Thus, different choices for $p_\perp$ may result in different numerical values for the zero points. 

\begin{table}
\begin{center}
\renewcommand\arraystretch{1.2}
\caption{\label{tab:2}The mean wavelength $\lambda_m$, the pivot wavelength $\lambda_p$, the zero points in the VEGAMAG and the AB photometric systems, and the $l_2$-norms of the parallel and the orthogonal components for the four passbands described in this work.}
\begin{tabular}{|l|c|c|c|c|} \hline
parameter & $G$ & $H_p$ & $B_T$ & $V_T$ \\ \hline
$\lambda_m$ [nm] & 661.25 & 540.19 & 419.68 & 529.68 \\
$\lambda_p$ [nm] & 646.78 & 532.74 & 418.91 & 528.34 \\
zp [VEGA] & 25.853 & 26.006 & 24.996 & 25.040 \\
zp [AB] & 25.974 & 25.975 & 24.905 & 25.000 \\
$||p_\parallel||_2$ & 16.8179 & 13.0333 & 7.0095 & 7.8246 \\
$||p_\perp||_2$ & 4.3186 & 2.0731 & 1.9359 & 3.1433 \\ \hline
\end{tabular}
\end{center}
\end{table}

\section{Influence of wavelength resolution \label{sec:wavelengthResolution}}
Up to now, we have considered the SPD $s(\lambda)$, without taking into account that any SPD is known to us only with a certain wavelength resolution. Thus, if $s^\prime(\lambda)$ is the ''true'' SPD, i.e. the SPD a hypothetical instrument with infinite spectral resolving power would observe, we only have available the SPD with a finite spectral resolution. Both are related via a variable kernel convolution, i.e.
\begin{equation}
s(\lambda) = \int\limits_0^\infty\, L(\lambda,\lambda^\prime) \cdot s^\prime(\lambda^\prime)\, {\rm d}\lambda^\prime \quad .
\end{equation}
Here, $L(\lambda,\lambda^\prime)$ is a {\it line spread function}, fulfilling the normalisation
\begin{equation}
\int\limits_0^\infty\, L(\lambda,\lambda^\prime)\, {\rm d}\lambda^\prime = 1 \quad .
\end{equation}
If $L$ does not change with wavelength $\lambda$, but its value only depends on the difference in wavelengths $\lambda$ and $\lambda^\prime$, one may write $L(\lambda,\lambda^\prime) \equiv L(\lambda-\lambda^\prime)$, which turns the problem into a normal convolution problem. For the synthetic photometry, computed from the finite-resolution SPD, we thus obtain
\begin{eqnarray}
c & = & \int\limits_{\lambda_0}^{\lambda_1}\, p(\lambda) \, \int\limits_0^\infty\, L(\lambda,\lambda^\prime) \cdot s^\prime(\lambda^\prime)\, {\rm d}\lambda^\prime \, {\rm d}\lambda \\
 & = & \int\limits_{0}^{\infty}\, s^\prime(\lambda^\prime) \, \int\limits_{\lambda_0}^{\lambda_1} \, p(\lambda) \cdot L(\lambda,\lambda^\prime) \, {\rm d}\lambda \, {\rm d}\lambda^\prime \quad .
\end{eqnarray}
Thus, we obtain a good approximation of the synthetic photometry with a finite-resolution SPD, i.e.
\begin{equation}
\langle \, p \, | \, s \, \rangle \approx \langle\, p \, | \, s^\prime\, \rangle \quad ,
\end{equation}
if the approximation
\begin{equation}
\int\limits_{\lambda_0}^{\lambda_1} \, p(\lambda) \cdot L(\lambda,\lambda^\prime) \, {\rm d}\lambda \approx p(\lambda^\prime) \label{eq:23}
\end{equation}
is valid. The line spread function $L$ has to be sufficiently narrow compared to the wavelength-dependent structures in the passband $p(\lambda)$ to be negligible. For a typical passband, this condition is easily met, as passbands tend to be rather smooth functions in wavelength. If the passband is separated into a parallel and an orthogonal component, as done in this work, both components will however not be smooth in wavelength anymore. The condition specified by Eq. (\ref{eq:23}) in this case has to be fulfilled for both the parallel and the orthogonal component individually. One has therefore to make sure that the SPD for which the synthetic photometry is to be computed has a significantly higher spectral resolution than $p_\parallel(\lambda)$ and $p_\perp(\lambda)$. For the results presented in this work, the spectral resolving power ranges up to $\sim$ 1000. If the wavelength resolution of an SPD for which synthetic photometry is to be computed is lower than 1000, no accurate computation of the parallel and orthogonal contribution of the passband is possible, and only the full passband $p(\lambda)$ should be used.

\section{Summary and Discussion \label{sec:Summary}}

We have discussed the problem of deriving a passband from photometric measurements and SPDs of a set of calibration sources. It was demonstrated that this problem has no unique solution. The shape of the passband remains widely unconstrained, with the uncertainty depending on the set of calibration SPDs. The passband can however be described by a sum of two components. One of these components (denoted the {\it parallel component}) is uniquely determined by the set of calibration SPDs. It can be determined by solving a system of linear equations. The other component (denoted the {\it orthogonal component}) is fully unconstrained and has to be estimated. Methods for deriving an estimate that satisfies the physical constraints on the passband (that is, smoothness and non-negativity) and that includes a-priori knowledge on the passband (e.g. previous laboratory measurements or simulations) has been presented.\par
The approach to passband determination as outlined in this work provides two advantages as compared to the previously used technique of modifying an initial passband guess until good agreement between observed and synthetical photometry is achieved. First, by formulating the intrinsically linear problem of passband determination as a set of linear equations, it can be ensured that the optimal solution for the passband (with respect to the set of calibration sources) is actually found. If an initial passband guess is modified in some way, without knowledge on the constrained component, it is a-priori impossible to know whether the choice for the modification method is actually able to represent a passband which contains the optimal solution. If this is not possible, one may end the process of passband determination with a sub-optimal solution without noticing. Second, the approach lined out in this work separates the constrained and the unconstrained elements in the process of passband determination. This allows to identify SPDs for which the synthetic photometry depends significantly on the unconstrained component of the passband, and which thus is subject to systematic uncertainties. If the passband is determined by the simpler modification of the initial passband guess only, then, even if the solution represents the constrained component correctly, it remains still impossible to identify SPDs affected by the systematic uncertainty resulting from the guess on $p_\perp(\lambda)$ that is made only implicitly in this approach.\par
Based on the findings of this work, a quantitative strategy for the selection of calibration sources for passband determination can be selected. Calibration sources are optimally selected such that the SPDs that are of scientific priority can be well expressed by a linear combination of the SPDs of the calibration sources. Such a selection minimises the dependency of the synthetic photometry of SPDs of interest on the unconstrained parallel component of the passband.\par
It should be stressed that including more calibration sources of the same spectral type can improve the determination of the parallel component of the passband, but it cannot compensate the lack of calibration sources with SPDs which are not linear combinations of the SPDs of the already used calibrations sources. Only the use of additional calibration sources with SPDs which increase the dimensionality of the sub-space of ${\mathcal L}^2$ spanned by the SPDs of calibrations sources can reduce the degeneracy in passband determination. An example for this problem is the discussed case of M$-$type stars. Calibration sources of M$-$type are required to achieve a reliable synthetic photometry of such cool objects.\par
In practice, the criterion for optimal choice of the calibration sources can hardly be fulfilled completely. It is therefore of interest to provide a simple measure on how sensitive the synthetic photometry of some SPD is to the estimation of the unconstrained component. To provide a simple indicator, we propose not only to use the solution for the passband, but both the parallel and orthogonal component of the passband. A simple measure (such as the angle $\gamma$) can then be obtained which indicates the relevance of the parallel component for any SPD.\par
We furthermore demonstrated that for photometric measurements of high precision, such as HIPPARCOS or {\it Gaia}, the uncertainty in the SPDs of the calibration stars dominate the uncertainty in the passband. The uncertainty in the SPDs includes also uncertainties in its shape, which may not be negligible compared to the uncertainty in absolute flux level. We suggest to quantify the uncertainty in shape by applying functional principal component analysis to a number different measurements of a SPD. By doing so, the SPD can be represented by a mean function, and the uncertainties on that mean curve by a small number of functions, weighted by uncorrelated random numbers. This representation is compact and can be included easily in error propagation. However, we found that the currently available information on the uncertainties in the shape of SPDs is not sufficient to perform a stringent error computation in this work.\par
We applied the methods developed in this work to the passbands of HIPPARCOS, Tycho, and {\it Gaia} DR1. We found equivalent (for $H_p$) or slightly better (for $B_T$, $V_T$, and $G$) passband solutions as compared to previous publications, as far as the level of relative flux residuals of the calibration sources used in this work is concerned. The differences in the level of relative flux residuals between the passbands derived in this work and other publications however are, with the exception of the $H_p$ passband by \citet{ESA1997} and the pre-launch estimate for the $G$ passband, so small, that they lie within the limits of uncertainty introduced by the calibration sources of this work.

\begin{acknowledgements}
      This work was supported by the MINECO (Spanish Ministry of Economy) through grants ESP2016-80079-C2-1-R (MINECO/FEDER, UE) and ESP2014-55996-C2-1-R (MINECO/FEDER, UE) and MDM-2014-0369 of ICCUB (Unidad de Excelencia ''Mar{\'i}a de Maeztu'').\\
      This work has made use of data from the European Space Agency (ESA) mission {\it Gaia} (\url{https://www.cosmos.esa.int/gaia}), processed by the {\it Gaia} Data Processing and Analysis Consortium (DPAC, \url{https://www.cosmos.esa.int/web/gaia/dpac/consortium}). Funding for the DPAC has been provided by national institutions, in particular the institutions participating in the {\it Gaia} Multilateral Agreement.
\end{acknowledgements}

%   \bibliography{AandA} % your references Yourfile.bib

\begin{thebibliography}{}

   \bibitem[Bessell (2000)]{Bessell2000} Bessell, M. 2000, PASP, 112, 961

   \bibitem[Bessell \& Murphy (2012)]{BessellMurphy} Bessell, M., \& Murphy, S. 2012,
      PASP, 124, 140

   \bibitem[Bohlin et al. (2017)]{Bohlin2017} Bohlin, R. C., M{\'e}sz{\'a}ros, S., Fleming, S. W., et al. 2017,
   AJ, 153, 234

  \bibitem[Carrasco et al. (2016)]{Carrasco2016} Carrasco, J. M., Evans, D. W., Montegriffo, P., et al. 2016,
  A\&A, 595, A7

   \bibitem[Chen et al. (2014)]{XSL} Chen, Y.-P., Trager, S. C., Peletier, R. F., Lan\c{c}on, A., Vazdekis, A., Prugniel, Ph., Silva, D. R., \& Gonneau, A. 2014,
   A\&A, 565, A117
   
   \bibitem[Coelho (2014)]{Coelho2014} Coelho, P.~R.~T. 2014,
   MNRAS, 440, 1027   

   \bibitem[ESA (1997)]{ESA1997} ESA 1997,
   The Hipparcos and Tycho Catalogues,
   ESA SP 1200
   
   \bibitem[Falc{\'o}n-Barroso et al. (2011)]{MILES} Falc{\'o}n-Barroso, J., S{\'a}nchez-Bl{\'a}zquez, P., Vazdekis, A., et al. 2011,
   A\&A, 532, A95
   
   \bibitem[Gaia Collaboration, Prusti  et al. (2016)]{Gaia2016a} Gaia Collaboration, Prusti, T., de Bruijne, J. H. J., Brown, A. G. A., et al. 2016,
   A\&A, 595, A1

   \bibitem[Gaia Collaboration, Brown et al. (2016)]{Gaia2016b} Gaia Collaboration, Brown, A. G. A., Vallenari, A., Prusti, T., et al. 2016,
   A\&A, 595, A2

   \bibitem[Heap \& Lindler (2016)]{HeapLindler} Heap, S. R., \& Lindler, D. 2016,
      in ASP Conference Series, Vol. 503,
      Calibration and Standardization of Missions and Large Surveys in Astronomy and Astrophysics, eds. S. Deustua, A. Allam, D. Tucker, \& J. A. Smith

   \bibitem[H{\o}g et al. (2000)]{Hoeg2000} H{\o}g, E., Fabricius, C., Makarov, V. V., Urban, S., Corbin, T., Wycoff, G., Bastian, U., Schwekendiek, P., \& Wicenec, A. 2000,
   A\&A, 355, L27

  \bibitem[Johnson (1952)]{Johnson1952} Johnson, H. L. 1952,
  ApJ, 116, 272

   \bibitem[Jordi et al. (2010)]{Jordi2010} Jordi, C., Gebran, M., Carrasco, J. M., de Bruijne, J., Voss, H., Fabricius, C., Vallenari, A., Kohley, R., \& Mora, A. 2010,
   A\&A, 523, A48

   \bibitem[Koornneef et al. (1986)]{Koornneef} Koornneef, J., Bohlin, R., Buser, R., Horne, K., \& Turnshek, D. 1986,
   in Highlights of Astronomy, ed. J.-P. Swings

   \bibitem[Le Borgne et al. (2003)]{Stelib} Le Borgne, J.-F., Bruzual, G., Pell{\'o}, R., Lan\c{c}on, A., Rocca-Volmerange, B., Sanahuja, B., Schaerer, D., Soubiran, C., \& V{\'i}lchez-G{\'o}mez, R. 2003,
   A\&A, 402, 433

   \bibitem[Lejeune et al. (1997)]{BaSeL} Lejeune, T., Cuisinier, F., \& Buser, R. 1997,
   A\&A Suppl. Ser., 125, 229

  \bibitem[Ma{\'i}z Apell{\'a}niz (2017)]{Maiz2017} Ma{\'i}z Apell{\'a}niz, J. 2017,
  A\&A, 608, L8

   \bibitem[Mann \& von Braun (2015)]{MannvonBraun2015} Mann, A. W. \& von Braun, K. 2015,
   PASP, 127, 102

   \bibitem[Markovsky et al. (2006)]{Markovsky2006} Markovsky, I., Rastello, M. L., Premoli, A., Kukush, A., \& van Huffel, S. 2006,
   Computational Statistics \& Data Analysis, 50, 181
   
   \bibitem[Milne (1980)]{Milne1980} Milne, R. D. 1980,
   Applied functional analysis: an introductory treatment, Pitman Advanced Pub. Program

   \bibitem[Oke \& Gunn (1983)]{OkeGunn} Oke, J. B. \& Gunn, J. E. 1983,
   ApJ, 266, 713

  \bibitem[Pancino et al. (2012)]{Elena2012} Pancino, E. Altavilla, G., Marinono, S., et al. 2012,
  MNRAS, 426, 1767

   \bibitem[Perryman et al. (1997)]{Perryman1997} Perryman, M. A. C., Lindegren, L., Kovalevsky, J., et al. 1997,
   A\&A, 323, L49
   
   \bibitem[Ramsay \& Silverman (2006)]{Ramsay2006} Ramsay, J. O. \& Silverman, B. W. 2006,
   Functional Data Analysis, Springer Series in Statistics, 2nd Edition
   
   \bibitem[Straizys \& Sviderskiene (1972)]{Vilnius} Straizys, V. \& Sviderskiene, Z. 1972,
   Astron. Obs. Bull. Vilnius, 35, 1
   
   \bibitem[Valdes et al. (2004)]{Valdes2004} Valdes, F., Gupta, R., Rose, J. A., Singh, H. P., \& Bell, D. J. 2004,
   ApJS, 152, 251
   
   \bibitem[van Leeuwen et al. (2017)]{vanLeeuwen2017} van Leeuwen, F., Evans, D. W., De Angeli, F., et al. 2017,
   A\&A, 599, A32
   
   \bibitem[Yao et al. (2005)]{Yao2005} Yao, F.,  Mueller, H.-G., \& Wang, J.-L.
      2005, Journal of the American Statistical Association, 100, 577
      
   \bibitem[Young (1994)]{Young1994} Young, A. T. 1994,
   A\&A, 288, 683
   
   \bibitem[Zeidler (1995)]{Zeidler1995} Zeidler, E. 1995,
   Applied functional analysis: applications to mathematical physics, Springer applied mathematics series 108
      
\end{thebibliography}

\end{document}